\documentclass[aps,prb,reprint,superscriptaddress,onecolumn,showpacs,notitlepage]{revtex4-2}
\setcitestyle{numbers,square}

\usepackage[T1]{fontenc}
\usepackage[utf8]{inputenc}
\usepackage{textcomp}
\usepackage{amsmath}
\usepackage{mathtools,amssymb,booktabs}

\usepackage{gensymb}

\usepackage{graphicx}

\usepackage{hyperref}
\hypersetup{
  colorlinks=true,
  linkcolor	=black,
  citecolor	=blue,
  urlcolor	=blue
}

\begin{document}

\title{Investigating the Electronic and Magnetic Properties of Na$_x$Fe$_{1/2}$Mn$_{1/2}$O$_2$
Cathode Materials with X-ray Compton Scattering}

\author{Veenavee Nipunika Kothalawala}
\email{veenavee.kothalawala@lut.fi}
\affiliation{Department of Physics, School of Engineering Sciences, LUT University, FI-53851 Lappeenranta, Finland.}

\author{Kosuke Suzuki}
\affiliation{Graduate School of Science and Technology, Gunma University, Kiryu, Gunma 376-8515, Japan. }

\author{Johannes Nokelainen}
\affiliation{Department of Physics, School of Engineering Sciences, LUT University, FI-53851 Lappeenranta, Finland.}
\affiliation{Department of Mechanical Engineering, School of Energy Systems, LUT University, FI-53851 Lappeenranta, Finland. }
\affiliation{Department of Physics, Northeastern University, Boston, Massachusetts  02115, USA. }
\affiliation{Quantum Materials and Sensing Institute, Northeastern University, Burlington, MA 01803, USA. }

\author{Ilja Makkonen}
\affiliation{Department of Physics, University of Helsinki, P.O. Box 43, FI-00014 University of Helsinki, Helsinki, Finland. }

\author{Erica West}
\affiliation{Department of Computational Engineering, School of Engineering Sciences, LUT University, FI-53851 Lappeenranta, Finland. }

\author{Lassi Roininen}
\affiliation{Department of Computational Engineering, School of Engineering Sciences, LUT University, FI-53851 Lappeenranta, Finland. }

\author{Jere Leinonen}
\affiliation{Research Unit of Sustainable Chemistry, University of Oulu, Oulu, Finland. }
\affiliation{Kokkola University Consortium Chydenius, University of Jyväskylä, Kokkola, Finland. }

\author{Pekka Tynjälä}
\affiliation{Research Unit of Sustainable Chemistry, University of Oulu, Oulu, Finland. }
\affiliation{Kokkola University Consortium Chydenius, University of Jyväskylä, Kokkola, Finland. }

\author{Petteri Laine}
\affiliation{Research Unit of Sustainable Chemistry, University of Oulu, Oulu, Finland. }
\affiliation{Kokkola University Consortium Chydenius, University of Jyväskylä, Kokkola, Finland. }

\author{Juho Välikangas}
\affiliation{Research Unit of Sustainable Chemistry, University of Oulu, Oulu, Finland. }
\affiliation{Kokkola University Consortium Chydenius, University of Jyväskylä, Kokkola, Finland. }

\author{Ulla Lassi}
\affiliation{Research Unit of Sustainable Chemistry, University of Oulu, Oulu, Finland. }
\affiliation{Kokkola University Consortium Chydenius, University of Jyväskylä, Kokkola, Finland. }

\author{Assa Aravindh Sasikala Devi}
\email{assa.sasikaladevi@oulu.fi}
\affiliation{Research Unit of Sustainable Chemistry, University of Oulu, Oulu, Finland. }
\affiliation{Materials and Mechanical Engineering Research Unit, University of Oulu, Oulu, Finland. }

\author{Matti Alatalo}
\affiliation{Nano and Molecular Systems Research Unit,  University of Oulu, Pentti Kaiteran Katu 1,  90570 Oulu, Finland. }

\author{Yuki Mizuno}
\affiliation{Japan Synchrotron Radiation Research Institute (JASRI), Sayo, Hyogo 679-5198, Japan. }

\author{Naruki Tsuji}
\affiliation{Japan Synchrotron Radiation Research Institute (JASRI), Sayo, Hyogo 679-5198, Japan. }

\author{Hikaru Usami}
\affiliation{Graduate School of Science and Technology, Gunma University, Kiryu, Gunma 376-8515, Japan. }

\author{Yuju Nagasaki}
\affiliation{Graduate School of Science and Technology, Gunma University, Kiryu, Gunma 376-8515, Japan. }

\author{Tsuyoshi Takami}
\affiliation{Otemon Gakuin University, 2-1-15 Nishiai, Ibaraki, Osaka 567-8502, Japan. }

\author{Yoshiharu Sakurai}
\affiliation{Japan Synchrotron Radiation Research Institute (JASRI), Sayo, Hyogo 679-5198, Japan. }

\author{Hiroshi Sakurai}
\affiliation{Graduate School of Science and Technology, Gunma University, Kiryu, Gunma 376-8515, Japan. }

\author{Mohammad Babar}
\affiliation{College of Engineering, Aerospace Engineering, University of Michigan, Ann Arbor, MI 48109 USA. }

\author{Venkat Vishwanathan}
\affiliation{College of Engineering, Aerospace Engineering, University of Michigan, Ann Arbor, MI 48109 USA. }

\author{Arun Bansil}
\affiliation{Department of Physics, Northeastern University, Boston, Massachusetts  02115, USA. }
\affiliation{Quantum Materials and Sensing Institute, Northeastern University, Burlington, MA 01803, USA. }

\author{Bernardo Barbiellini}
\affiliation{Department of Physics, School of Engineering Sciences, LUT University, FI-53851 Lappeenranta, Finland.}
\affiliation{Department of Physics, Northeastern University, Boston, Massachusetts  02115, USA. }
\affiliation{Quantum Materials and Sensing Institute, Northeastern University, Burlington, MA 01803, USA. }

\date{\today}

\begin{abstract}
  We discuss electronic and magnetic properties of Na$_x$Fe$_{1/2}$Mn$_{1/2}$O$_2$,
  a promising Na-ion battery cathode material. Using x-ray Compton scattering, SQUID magnetometry, and density-functional-theory based modeling, we probe how electrons and spins evolve during sodiation. By comparing Compton profiles of sodiated and desodiated samples, we show that oxygen 2$p$ orbitals drive the redox process, while transition-metal 3$d$ electrons become more delocalized, explaining the metallic phase at $x=2/3$. These profile differences define a quantitative descriptor for the sodiation range associated with improved conductivity. Electron holes on oxygen, reflected in oxygen magnetization,
  confirm the important role of oxygen in the electrochemical activity of the cathode.
\end{abstract}

\maketitle

\section{Introduction}

Uneven geographical distribution of lithium (Li) reserves has significant societal impacts~\cite{Elsa2017,quinteros2020} in terms of the sustainability of lithium-ion batteries for grid-scale energy storage systems. As a result, sodium-ion batteries (NIBs) have been gaining increasing attention due to their lower cost and the greater abundance of sodium (Na). The first commercially available NIBs were offered by Asian manufacturers and have been the subject of extensive analyses~\cite{dorau2024}. Nevertheless, NIBs face challenges due to the larger atomic radius of Na compared to Li ions, resulting in slower Na$^+$ diffusion kinetics that leads to lower energy density and shorter cycle life.

Performance of NIBs is primarily influenced by the choice of cathode material. Among the various candidate materials, layered oxides, particularly NaMnO$_2$, has shown great potential for applications~\cite{jamil2023,rostami2024}. NaMnO$_2$ based materials can be synthesized in different structural phases, such as O2, P2, O3 and P3, depending on the arrangement of the oxide layers and the sodium environment~\cite{delmas19801}: here, `O' denotes an octahedral environment for Na ions, while `P' indicates a prismatic environment, and the number in O2, P2, O3 and P3 refers to the number of distinct interlayers involved in various  oxide layer packings. Phase transitions between these structures can occur during the charging process, leading to cathode degradation.

\begin{figure}
  \centering
  \includegraphics[width=0.45\linewidth]{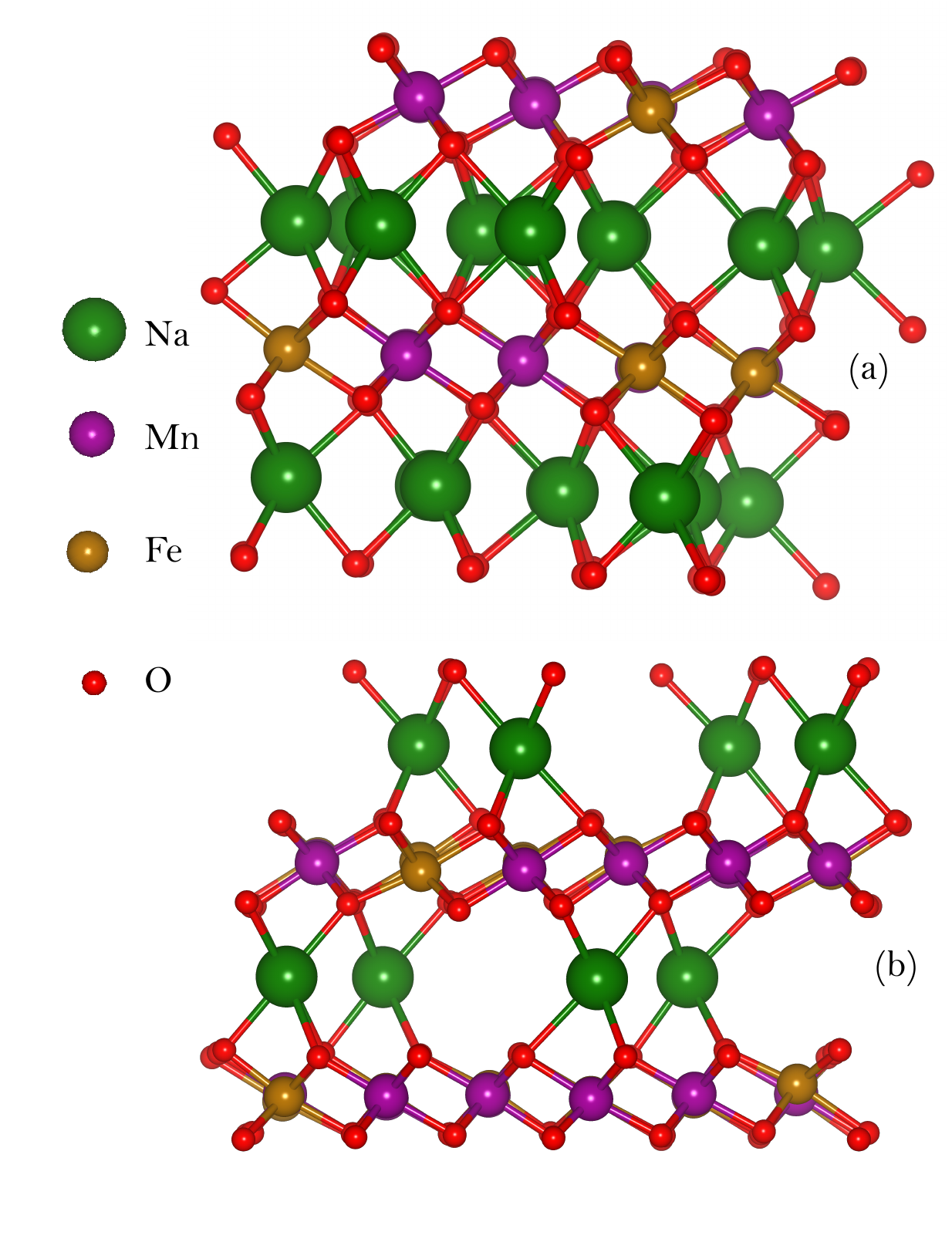}
    \caption{Structures of (a) P2-Na$_{2/3}$Fe$_{1/2}$Mn$_{1/2}$O$_2$ and (b) O2-Na$_{2/9}$Fe$_{1/2}$Mn$_{1/2}$O$_2$ used in our DFT calculations.}
   \label{Fig:structure}
\end{figure}

Recent studies have focused on addressing these challenges by substituting transition metals in the cathode structure. For example, doping with iron (Fe) has been shown to mitigate structural degradation and improve cycling performance.
This strategy has led to the development of novel sodium-ion electrode materials,
such as P2-Na$_x$Fe$_{1/2}$Mn$_{1/2}$O$_2$, which exhibits promising reversible capacity and it is stable between $x \sim 0.4$ and $x = 2/3$ at voltages as high as $3.8$ V~\cite{yabuuchi2012}.
However, when charged to $4.2$ V, the diffraction patterns indicate a phase transition from the P2 to the O2 phase, which can lead to performance degradation~\cite{yabuuchi2012}.
The P2 and O2 structures are illustrated in
Fig.~\ref{Fig:structure}.
Interestingly, Tang {\it et al.}~\cite{tang2024} recently suggested that this phase transition issue could be addressed by introducing vacancies into the transition metal layer.
Moreover, Wang {\it et al.}~\cite{wang2025} show that oxygen vacancies (OVs) play a crucial role in enhancing Mn redox activity by lowering the Mn valence state. OVs, which are formed intrinsically or by anionic redox during high-voltage operation, also promote reversible Fe$^{3+}$ migration to stabilize the structure and support redox reactions involving the oxygen sublattice. This dynamic interplay strengthens charge compensation and improves electrochemical performance. Overall, controlled OV engineering helps mitigate Mn$^{3+}$ Jahn-Teller distortion and supports the development of stable, high-capacity Fe‚Mn layered cathodes for NIBs.

To facilitate further advances, a comprehensive, atomic-level understanding of the fundamental processes underlying the redox reactions in the NIBs is necessary. In this way, rational strategies for optimizing the performance and durability of sodium-based batteries can be designed.

Here we use both regular and magnetic Compton x-ray scattering experiments combined with first-principles density functional theory (DFT) modeling to investigate the layered oxide cathode material Na$_x$Fe$_{1/2}$Mn$_{1/2}$O$_2$.
By measuring Compton profiles at sodium concentrations of $ x = \frac{2}{3} $ and $x = \frac{1}{3} $, we extract detailed information about the redox orbitals involved in the battery's operation. Moreover, we identify the magnetic orbitals associated with these two sodium concentrations by analyzing the associated the magnetic Compton profiles.
We emphasize that (non-magnetic) difference Compton profiles and the magnetic Compton profiles yield complementary insights into the redox and magnetic orbitals, respectively~\cite{kothalawala2024a}.

While Compton scattering has historically been viewed as a specialized technique, its scope has broadened significantly over the years.
It is a nondestructive bulk-sensitive probe that measures the ground-state electronic momentum density of materials. Unlike conventional spectroscopies that are surface-sensitive or rely on optical transitions, Compton scattering directly probes the spatial distribution of valence electrons in momentum space. This enables the detection of changes in orbital occupancy and electron delocalization that accompany redox processes or magnetic ordering. In the context of battery materials, X-ray Compton scattering provides unique insights into the spatial extent and nature of redox orbitals, distinguishing between localized and delocalized characters, and thereby informing mechanisms of phase stability, capacity retention, and degradation as demonstrated in LiCoO$_2$ cathode materials~\cite{barbiellini2016,nokelainen2022}. A recent review by Zuo {\em et al.}~\cite{khalil2025} shows that Compton scattering can be applied to characterize commercial batteries under \emph{operando} conditions, providing valuable information on Li-ion flow, degradation pathways, and structural changes during cycling. These capabilities position Compton scattering as a complementary tool to the more conventional spectroscopic and diffraction methods, and as a way of identifying spectroscopic descriptors that provide atomic-scale insight into the redox mechanisms underlying the operation of Li-ion batteries~\cite{suzuki2021,Suzuki2024}.
Given the current interest in electrochemical energy storage, the integration of Compton scattering into this domain can be expected to provide new opportunities to connect fundamental atomic-level understanding of materials with macroscopic device performance.

\subsection{Samples}
Fe$_{1/2}$Mn$_{1/2}$CO$_3$ precursors were synthesized by carbonate coprecipitation in an inert nitrogen atmosphere to prevent oxidation. The coprecipitation was carried out in a stirred-tank reactor using an overhead stirrer and a heating plate. An Aqueous solution of metal sulfates (2~M FeSO$_4$ + MnSO$_4$) and 2~M Na$_2$CO$_3$ was fed into the reactor using peristaltic pumps. The feed rate of the metal solution was kept constant, while the pH was maintained at approximately 8 by adjusting the feed rate of the Na$_2$CO$_3$ solution. The temperature inside the reactor was held at 50$^{\circ}$C throughout the 6h coprecipitation process. Similar coprecipitation conditions have been reported to strongly influence particle size distribution and precursor properties in Fe/Mn carbonate systems~\cite{Jere2025}.

After coprecipitation, the precursor slurry was vacuum-filtered, and the precipitate was thoroughly washed with deionized water. The resulting Fe$_{1/2}$Mn$_{1/2}$CO$_3$ precursors were dried overnight at 60$^{\circ}$C in a vacuum oven. The particle size distribution of the precursor powder, measured by laser diffraction, yielded characteristic diameters of D$_1$ = 6.75~$\mu$m, D$_{10}$ = 8.11~$\mu$m, D$_{50}$ = 10.6~$\mu$m, D$_{90}$ = 13.6~$\mu$m, and D$_{99}$ = 15.9\,$\mu$m.
The dried precursor material was subsequently milled and sieved under dry-room conditions. It was mixed with anhydrous Na$_2$CO$_3$ (99.5\%, Alfa Aesar) using an IKA A11 basic mixer with two mixing cycles of 30~s separated by a cooling interval. Prior to mixing, the Na$_2$CO$_3$ was ground and sieved using a mortar and pestle. The powder mixtures were placed in alumina crucibles and calcined at 900$^{\circ}$C for 12~h under air, using a heating ramp of 2.5$^{\circ}$C/min. Molar ratios of Na:(Fe$_{0.5}$Mn$_{0.5}$CO$_3$) = 2/3 and 1/3 were used to synthesize Na$_{2/3}$Fe$_{1/2}$Mn$_{1/2}$O$_2$ and Na$_{1/3}$Fe$_{1/2}$Mn$_{1/2}$O$_2$, respectively. The small amount of sodium impurity present in the precursors was taken into account during weighing. After cooling, the calcined products were ground using a mortar and pestle and sieved through a 25~$\mu$m aperture sieve.

The crystal structures of the synthesized materials were determined by X-ray diffraction (XRD) using a SmartLab diffractometer (Rigaku Corporation) with CuK$_{\alpha}$ radiation (see Supplemental Materials). The sample with $x = 2/3$ crystallizes in the P2 phase. The $x = 1/3$ sample also adopts the P2 phase, despite being located at the lower boundary of the reported P2 phase stability range ($x \sim 0.4$ to $x = 2/3$)~\cite{yabuuchi2012}, although it exhibits additional weak diffraction peaks indicative of minor impurity phases, as shown in Fig.~S2 of the Supplemental Materials.

Magnetization measurements were performed using a SQUID magnetometer (MPMS-7, Quantum Design, Inc.). Measurements were conducted at 10~K by sweeping the magnetic field from $-2.5$\,T to 2.5\,T.

\subsection{DFT Based Modeling}

Our DFT calculations were based on the projector-augmented-wave method~\cite{Blochl1994} as implemented in the Vienna \emph{ab initio} simulation package\cite{Kresse1996, Kresse1999}.
To account for strong electron correlation effects, we used the regularized strongly-constrained-and-appropriately-normed r$^2$SCAN~\cite{Furness2020} exchange-correlation functional which avoids the use of effective Hubbard parameters~\cite{dudarev1998}.
We employed a plane-wave kinetic energy cut-off of 480\,eV, which we found to provide a reasonable balance between computational cost and accuracy.
This choice is supported by a previous study on a related sodium-ion battery material by Lee {\it et al.}~\cite{shirley2013}, where convergence within approximately 3\,meV per formula unit was achieved using a 450\,eV cut-off, indicating that this value is adequate for the comparative purposes of our study.
Gaussian smearing with a width of 0.05\,eV (full width at half maximum) and a total energy tolerance of $10^{-6}$\,eV were used to determine the self-consistent charge density.
The Na ($2p$, $3s$), Fe ($3d$, $4s$), Mn ($3p$, $3d$, $4s$), and O ($2s$, $2p$) electrons were treated as valence electrons.

We based our DFT computations on atomic coordinates for the P2 ($x = 2/3$) and O2 ($x = 2/9$) phase models obtained by Zarrabeitia {\it et al.} in their detailed structural work~\cite{zarrabeitia2022}, where the Fe and Mn cations are randomly distributed within the transition-metal layers in a 1:1 ratio, see Fig.~\ref{Fig:structure}.
The P2-type structure corresponds to the sodiated phase formed at low voltage and the O2-type structure corresponds to the desodiated phase formed at high voltage.
We adopted a ferromagnetic high-spin configuration for Fe and Mn following Zarrabeitia {\it et al.}
A $4\times4\times1$ ($4\times2\times1$) $k$-mesh was used to sample the Brillouin zone of P2 (O2).
We relaxed these structures with an atomic force tolerance of 0.01 eV/\AA\,  and obtained the lattice constants of $a = 2.913\,\AA$, $c = 11.113\,\AA$ for the P2 and $a = 2.864\,\AA$, $c = 11.208\,\AA$ for the O2 phase.
The P2 phase values are close to the experimental values of $a= 2.9405$ \AA\  and $c = 11.1957$ \AA\ ~\cite{delmas2014}.
The Bader charge analysis~\cite{2009_Tang_Bader, 2011_Yu_Trinkle_Bader_improvement} was used to compute atomic charges based on the DFT calculations.

\subsection{Momentum Density Calculations}
As already pointed out above, Compton scattering with x-rays~\cite{Cooper2004,barbiellini2001} has proven particularly useful in the advanced characterization of batteries\cite{nokelainen2022,Suzuki2024}.
In a Compton scattering experiment, the electron momentum density $\rho({\bf p})$ is accessed via the Compton profile $J(p_z)$, which is derived from the Doppler broadening of the scattered photons. This analysis relies on the impulse approximation, which assumes that the binding energy of the electrons can be neglected. As a result, the Compton profile directly reflects the electronic momentum distribution, containing information about both localized and delocalized occupied orbitals. Kaplan {\it et al.}.~\cite{Kaplan2003} discuss the underlying many-body formalism based on Dyson orbitals. The impulse approximation holds well for high-energy ($\geq100$\,keV) x-ray photons. The Compton profile $J(p_z)$ is given by~\cite{Kaplan2003}:
\begin{equation}
J(p_z)= \iint \rho ({ p}) \mathrm{d}p_{x} \mathrm{d}p_{y}
\label{eq:one},
\end{equation}
where $p = (p_x, p_y, p_z)$ is the electron momentum,
and $\rho (p)$ is the electron momentum density,
which following Eq. (18) in {\it the Encyclopedia of Condensed Matter Physics}~\cite{barbiellini2024} can be expressed as
a sum over electronic orbitals as:
\begin{equation}
\rho({\bf p}) = \sum_{j} n_{j} \int \left\vert \Psi_{j}({\bf r}) \exp(-i{\bf p \cdot r})\,
\right\vert ^{2} \mathrm{d}^3{\bf r},
\label{eq:two}
\end{equation}
where $\Psi_{j}$ is a natural spin orbital and $n_{j}$ is the associated occupation number.
Electron momentum density $\rho({\bf p})$ is often approximated by
replacing the natural spin orbitals by the atomic Hartree-Fock orbitals~\cite{biggs1975} and Kohn-Sham Bloch orbitals~\cite{barbiellini2001}.
Within the independent particle model,
$n_{j} = 1$ if the state is occupied and $n_{j} = 0$ otherwise.
The integral of $J(p)$ gives the total number of electrons.

Magnetic Compton profile $\it{J}_{\text{mag}}(p_z)$ is given by
\cite{barbiellini2024}
\begin{equation}
J_{\text{mag}}\left(p_{z}\right)=\iint\left(\rho_{\uparrow}(\mathbf{p})-\rho_{\downarrow}(\mathbf{p})\right) \mathrm{d}p_{x} \mathrm{d}p_{y},
\label{eq:three}
\end{equation}
where $\rho_{\uparrow}(\mathbf{p})$ and $\rho_{\downarrow}(\mathbf{p})$ are the momentum densities of the majority and minority spins, respectively. Spin magnetic moment $\mu_\text{spin}$ is obtained by integrating $J_\text{mag}(p_z)$. The $z$-direction ($p_z$) here corresponds to the scattering vector, which is aligned with the photon momentum transfer direction. In our experimental geometry, this nearly coincides with the incident photon propagation axis. The magnetic field was applied along the same axis. Since our samples are polycrystalline with randomly oriented domains, the measured spin-up and spin-down momentum densities both become spherically averaged. To obtain spherical averages theoretically, we use a Monte Carlo sampling approach coupled with linear interpolation. Since our study focuses on spherically-averaged spectra, we will hereafter replace $p_z$ by the radial distance $p = \|\mathbf{p}\|$.

Usually, contributions to the Compton profiles from core orbitals\cite{suzuki2016} are taken from the tabulated Compton profiles obtained within the Hartree-Fock scheme~\cite{biggs1975}.
Our theoretical Compton profiles for the valence electrons are based on our DFT computations following the method of Makkonen {\em et al.}~\cite{Makkonen2005}.

\subsection{Compton Profile Measurements}
\begin{figure}
\centering
\includegraphics[width=.5\linewidth]{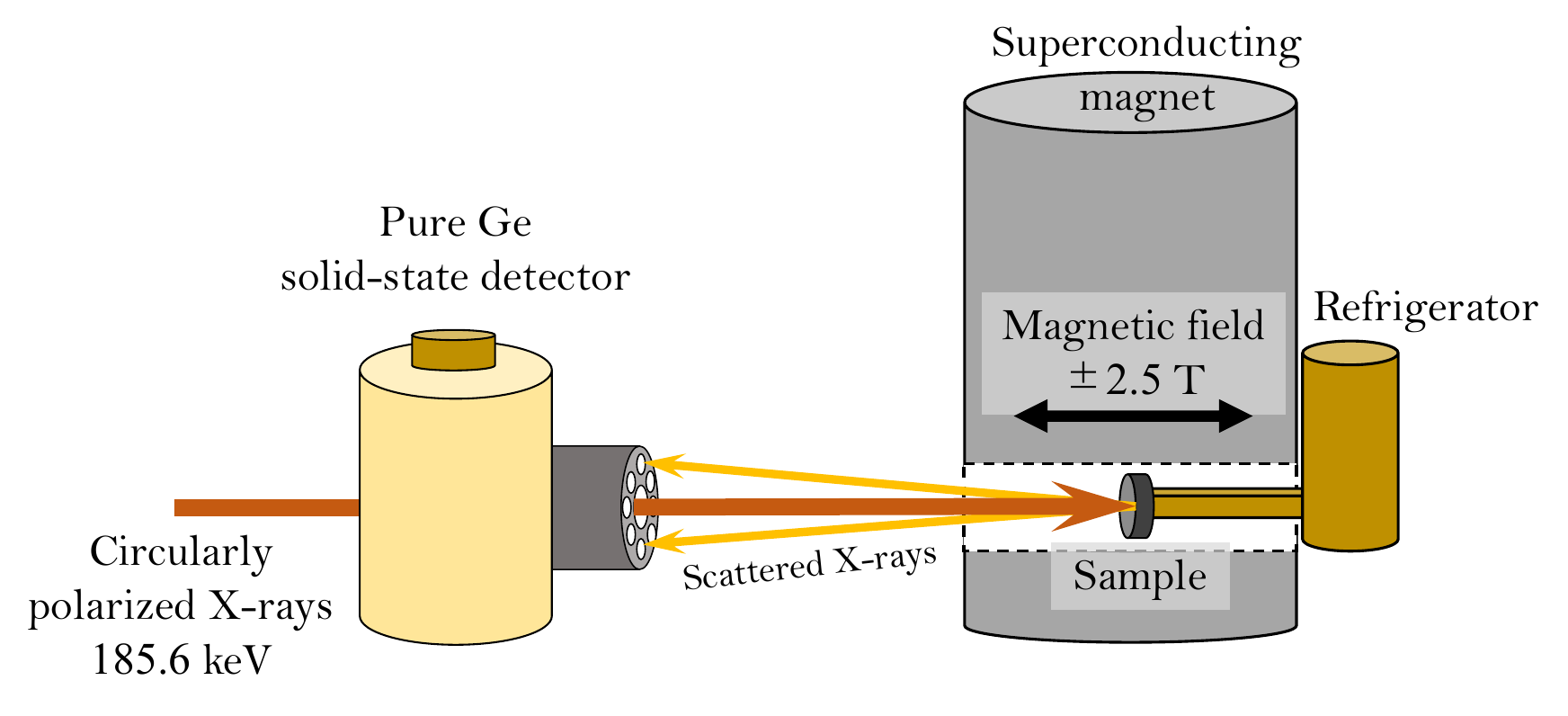}
\caption{
A schematic of the experimental magnetic Compton scattering setup at BL08W of SPring-8. Scattering geometry and orientation of the magnetic field are shown. The scattering vector lies along the z-direction of the Compton profile, which is nearly parallel to the incident photon direction, corresponding to a scattering angle of 178$^{\circ}$. Momentum densities are spherically averaged in polycrystalline samples, while the spin remains aligned with the magnetic field.}
\label{Fig:MCS}
\end{figure}

Non-magnetic and magnetic Compton profiles were measured at the high-energy inelastic scattering beamline 08W at the Japanese synchrotron facility SPring-8~\cite{Sakurai1998, Kakutani2003}.
The experimental setup is illustrated in Fig.~\ref{Fig:MCS}.
Circularly polarized x-rays of 182.6\,keV emitted from an elliptical multipole wiggler were irradiated to the sample. Size of the incident x-ray beam at the sample position was 1\,mm$^2$. Compton scattered x-rays were measured using a pure Ge solid-state detector. The scattering angle was fixed at 178 degrees. For the magnetic Compton profiles, a magnetic field of $\pm$ 2.5 \,T was applied to the sample that was kept at 7 \,K to obtain Compton scattered x-ray intensities, $I_+$ and $I_-$, by flipping the magnetic field every 60 \,s. Magnetic and non-magnetic Compton scattering experiments were both performed at 7 \,K.

\section{Results and Discussion}

 Bader charge analysis reveals that during the desodiation transition from the P2 (Na$_{2/3}$Fe$_{1/2}$Mn$_{1/2}$O$_2$) to the O2 (Na$_{2/9}$Fe$_{1/2}$Mn$_{1/2}$O$_2$) phase,
charge transfer mostly occurs at the oxygen sites.
Bader charges associated with oxygen atoms are on average $0.158\,e$ higher in the P2 compared to the O2 phase.
In contrast, the corresponding relative Bader charges on the transition metals are $0.131\,e$ for Fe and $0.022\,e$ for Mn.
$80.5\,\%$ of the additional charge on the Fe$_{1/2}$Mn$_{1/2}$O$_2$ planes in the P2 phase compared to the O2 phase is distributed on the oxygen sites.
Sodium atoms, which act primarily as charge compensators exhibit only minimal variation in charge ($0.020\,e$).
Bader analysis thus supports the conclusion that oxygen anions play the dominant role in charge compensation upon desodiation from P2 to O2.

\begin{figure}
 \centering
  \includegraphics[width=.5\linewidth]{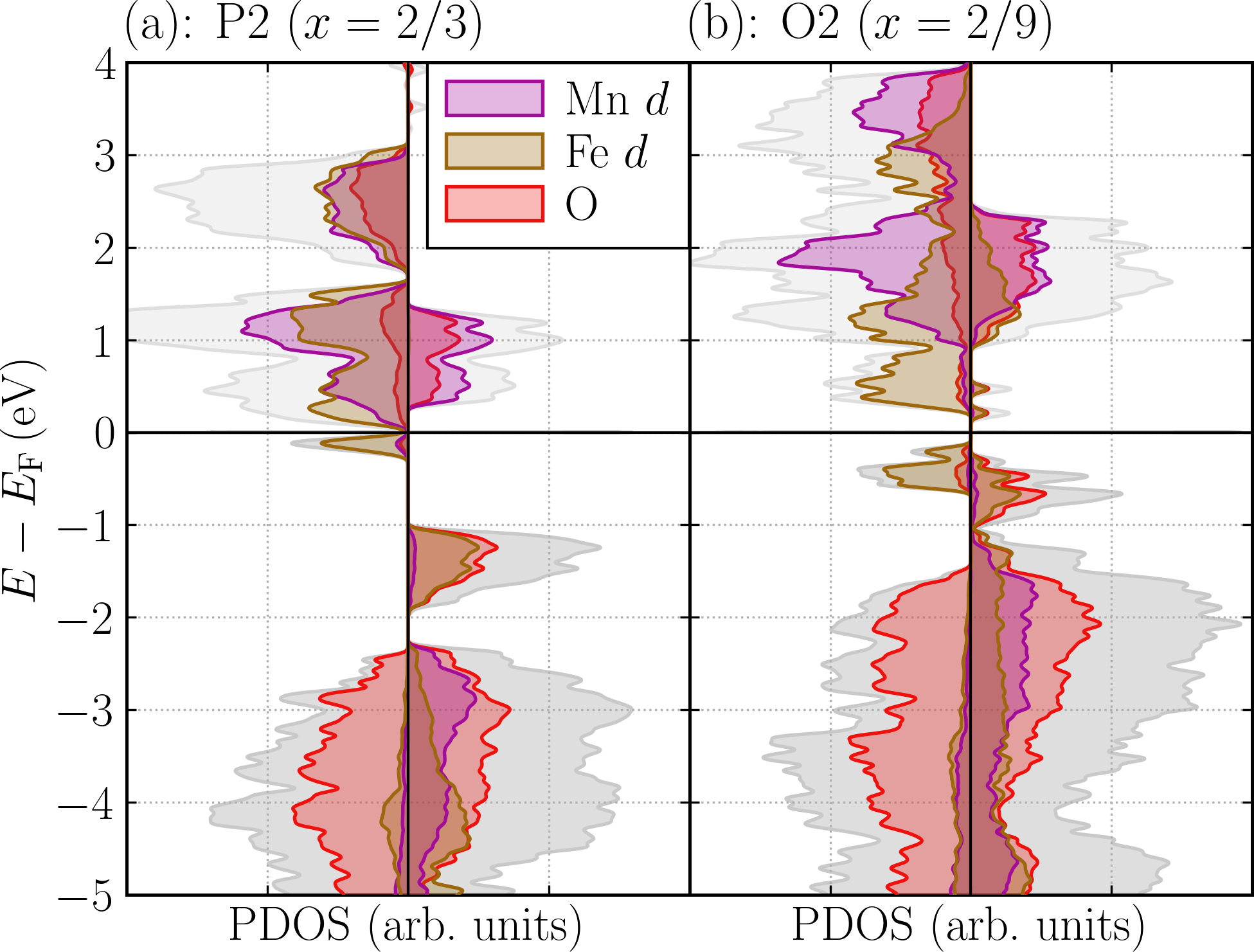}
  \caption{Spin and orbital resolved total and partial densities of states on various atomic sites (see legend) in (a) P2-Na$_{2/3}$Fe$_{1/2}$Mn$_{1/2}$O$_{2}$ and (b) O2-Na$_{2/9}$Fe$_{1/2}$Mn$_{1/2}$O$_{2}$.
  \label{Fig:DOS}}
\end{figure}

Spin-resolved partial DOS (PDOS) for P2-Na$_{2/3}$Fe$_{1/2}$Mn$_{1/2}$O$_2$ and O2-Na$_{2/9}$Fe$_{1/2}$Mn$_{1/2}$O$_2$ is shown in Fig.~\ref{Fig:DOS}.
Our PDOS for the P2-type structure does not exhibit a band gap between the valence and conduction bands but we observe a band gap for the O2-type structure.
This is in contrast to the results reported by Zarrabeitia \textit{et al.}\cite{zarrabeitia2022},
who used the PBE exchange-correlation functional~\cite{Perdew1996} with Hubbard corrections~\cite{dudarev1998} applied on the TM ions,
and observed band gap for P2 but not for O2.
Our results, however, are consistent with the measured electronic resistivity values, which show that the sodium-ion-extracted electrode possesses a more insulating character compared to the sodiated sample~\cite{zarrabeitia2022}. P2 PDOS results have also been reported by Abate \textit{et al.}\cite{abate2021} and Kim \textit{et al.}~\cite{kim2022}.
The DFT calculations of Abate \textit{et al.} do not display a gap near the Fermi energy, in agreement with our findings.
This is likely because these authors included a Hubbard correction on the oxygen atoms.
A consistent feature across all available PDOS results is the strong O-$2p$ character near the Fermi level, indicating a substantial oxygen contribution to the redox orbitals

\begin{figure}[ht]
 \includegraphics[width=.5\linewidth]
  {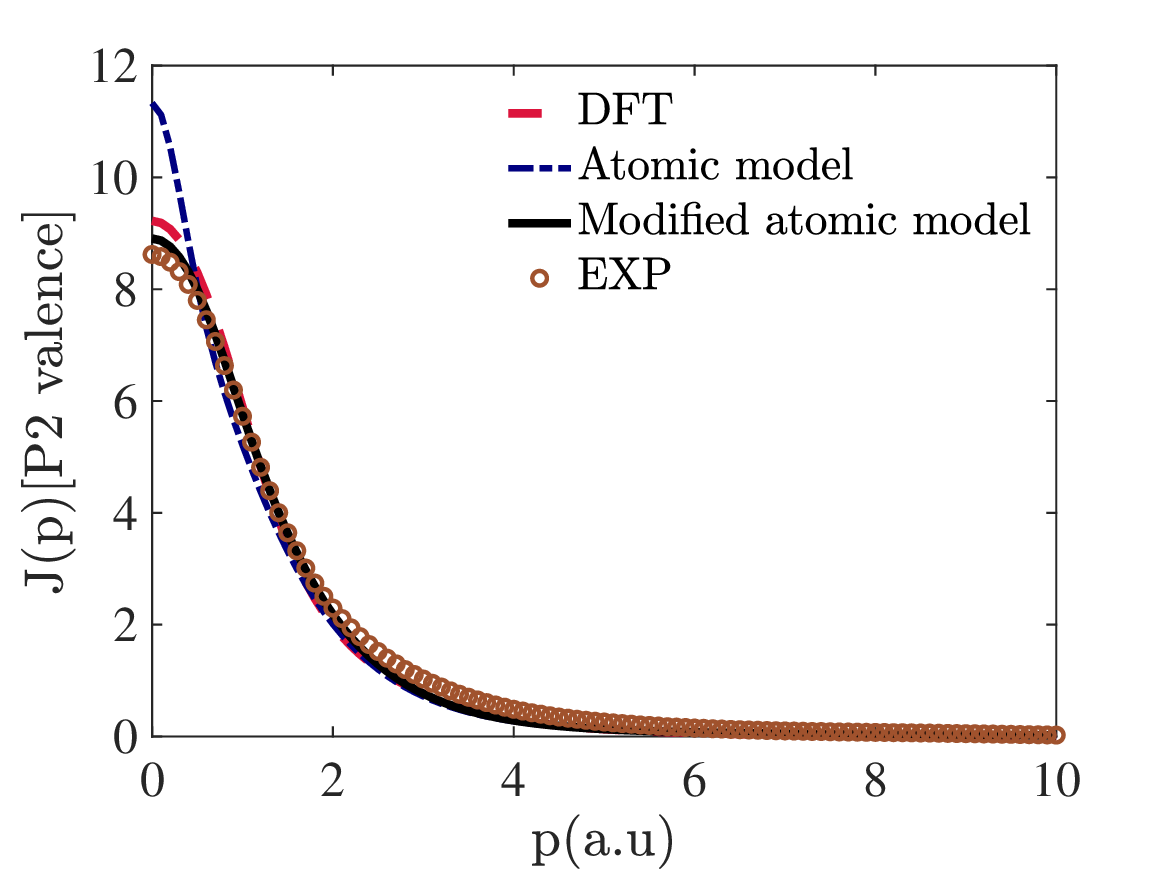}
   \caption{Spherically averaged theoretical and experimental valence Compton profiles of P$_2$-Na$_{2/3}$Fe$_{1/2}$Mn$_{1/2}$O$_{2}$. Results for three different theoretical models are given (see text for details). Theoretical profiles are convoluted with a Gaussian of 0.5 a.u. full-width-at-half-maximum.
\label{Fig:compton_profiles}}
\end{figure}

Experimental and computed valence Compton profiles of Na$_{2/3}$Fe$_{1/2}$Mn$_{1/2}$O$_2$ are presented in Fig.~\ref{Fig:compton_profiles}. We consider three different theoretical models. (1) An atomic orbital model~\cite{suzuki2016}. (2) A modified atomic orbital model~\cite{patel2022, kothalawala2024b}, in which all positive ions donate electrons to the O-$2p$ orbitals. And (3) first-principles DFT based results for the valence electrons.  The atomic orbital model is seen to be in poor agreement with the experimental profile, but the two other models are in better agreement. These results indicate that the electronic orbitals underlying these two models correctly capture the overall electron momentum distribution of this ionic layered oxide, and that the Na-$3s$ electron here is mostly donated to the O-$2p$ orbital.
Note that the modified atomic fit is consistent with ideal stoichiometry within the sensitivity of Compton scattering. In particular, oxygen vacancies at the level of several percent or higher would lead to a measurable redistribution of valence electrons, leaving excess charge on the Na-$3s$ states. This would produce a clear and systematic mismatch between the experimental and calculated Compton profiles, most prominently in the low-momentum region, which is highly sensitive to the characteristic momentum distributions of O-$2p$ and Na-$3s$ orbitals \cite{Kosuke2016}. The absence of such deviations therefore rules out a significant (at about the percent-level~\cite{kothalawala2024b}) concentration of defect vacancies in the present samples. An analysis along these lines for Na$_{1/3}$Fe$_{1/2}$Mn$_{1/2}$O$_2$ is presented in the Supplemental Materials (see Fig.~S3).

\begin{figure}
 \includegraphics[width=.5\linewidth]
  {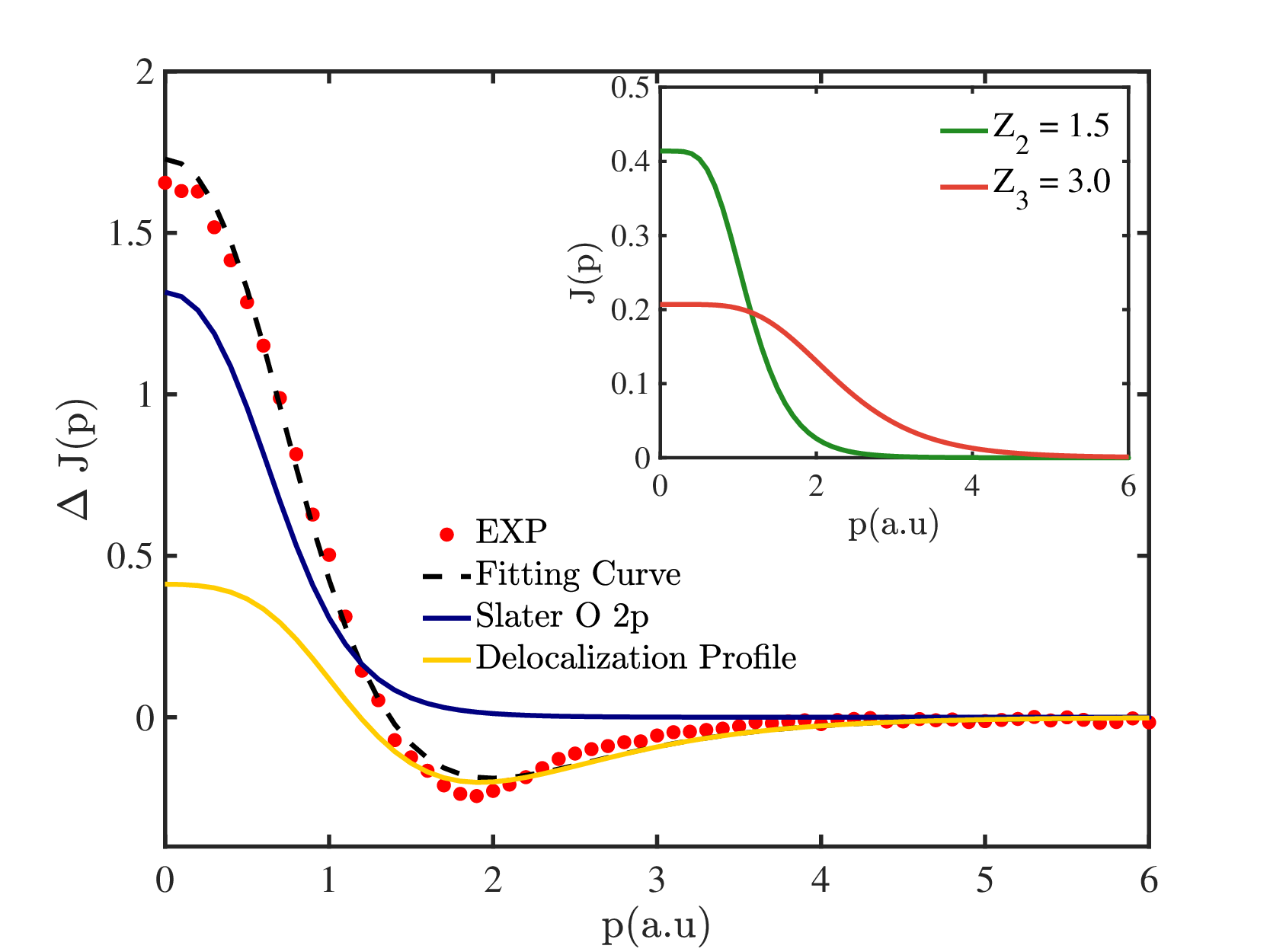}
   \caption{Experimental valence Compton profile difference $\Delta J(p)$ between P$_2$-Na$_{2/3}$Fe$_{1/2}$Mn$_{1/2}$O$_{2}$ and O$_2$-Na$_{1/3}$Fe$_{1/2}$Mn$_{1/2}$O$_{2}$, along with the corresponding curve-fitting results (see text for details), including contributions from O-$2p$ and Fe/Mn $3d$ electrons. Inset shows the Compton profiles of transition metal $3d$ orbitals for the indicated values of the $Z_2$ and $Z_3$ parameters. The profile $D(p)$ is obtained by taking the difference between the profiles for $Z_2 = 1.5$ and $Z_3 = 3$. The area under the negative excursion of the $D(p)$ profile quantifies the number of $3d$ electrons displaced during the sodiation process.}
\label{Fig:compton_profile_diff}
\end{figure}

To investigate the redox mechanism in detail, we adopt a method previously applied successfully to Li-ion battery materials~\cite{suzuki2015, barbiellini2016, hafiz2021, nokelainen2022, kothalawala2024a, Suzuki2024}. Specifically, we analyze the redox activity by subtracting the valence Compton profile of Na$_{2/3}$Fe$_{1/2}$Mn$_{1/2}$O$_2$ from that of Na$_{1/3}$Fe$_{1/2}$Mn$_{1/2}$O$_2$, thereby isolating the redox-induced changes in the electronic structure.
This difference profile, denoted as $\Delta J(p)$, is shown in Fig.~\ref{Fig:compton_profile_diff} and it is modeled using Slater-type orbitals~\cite{kothalawala2024a}, which capture key physical properties, such as the correct exponential decay of electron density with distance, given by
\begin{equation}
\psi(r) \propto r^{n-1} e^{-Z r},
\end{equation}
and a finite value at the nucleus.
The spherically-averaged Compton profile for an O-2$p$ orbital described by a Slater-type orbital is given by
\begin{equation}
\label{eq:Jp02p}
J_{2p}(Z_1,p) = \left( \frac{32Z_1^7}{15\pi} \right) \left( \frac{5p^2 + Z_1^2}{(p^2 + Z_1^2)^5} \right).
\end{equation}
This orbital represents the primary destination for redox electrons donated by sodium.
The spherical average of the Compton profile for a 3$d$ shell, modeled using cubic harmonics, can be expressed as
\begin{equation}
\label{eq:J3d}
\begin{split}
J_{3d}(Z,p) &= \frac{1}{5}\big[ J_{xy}(Z,p) + J_{x^2 - y^2}(Z,p) \\
&\quad + J_{yz}(Z,p) + J_{zx}(Z,p) + J_{z^2}(Z,p) \big],
\end{split}
\end{equation}
with the components given by:
\begin{align}
\label{eq:dxy}
J_{xy}(Z, p) &= J_{x^2 - y^2}(Z, p)
= \frac{128\, Z^9}{35\pi (Z^2 + p^2)^5}, \\[8pt]
\label{eq:dyz}
J_{yz}(Z, p) &= J_{zx}(Z, p)
= \frac{1280\, Z^9 p^2}{35\pi (Z^2 + p^2)^6}, \\[8pt]
\label{eq:dz2}
J_{z^2}(Z, p) &= \frac{256\, Z^9}{105\pi} \left[
    \frac{60\, p^4}{(Z^2 + p^2)^7}
  - \frac{10\, p^2}{(Z^2 + p^2)^6}
  + \frac{1}{(Z^2 + p^2)^5}
\right].
\end{align}
To investigate the redox process, we express the difference in valence Compton profiles between the two samples with different sodium concentrations as
\begin{equation}
\label{eq:deltaJ}
\Delta J(p) = J_{2p}(Z_1,p) + D(p),
\end{equation}
where $J_{2p}(Z_1, p)$ models the contribution of the electron transferred to the O 2$p$ orbital. $D(p)$ captures effects of reshuffling among the 3$d$ orbitals without a net occupancy change, and is given by:
\begin{equation}
\label{eq:D}
D(p) = J_{3d}(Z_2,p) - J_{3d}(Z_3,p),
\end{equation}
The best fit yields $Z_1 = 1$, $Z_2 = 1.5$, and $Z_3 = 3$. To quantify the number of displaced 3$d$ electrons, we integrate the absolute value of the redistribution term:
\begin{equation}
n_e = \frac{1}{2} \int dp\, |D(p)|,
\end{equation}
obtaining $n_e = 0.17$ electrons per Na atom. This quantitative insight—particularly the appearance of the negative excursion in $\Delta J(p)$—provides a useful metric for tracking 3$d$ electron delocalization and redox reversibility~\cite{barbiellini2016}, which are key factors governing the performance of battery cathodes~\cite{barbiellini2016}.

\begin{figure}
   \includegraphics[width=.5\linewidth]{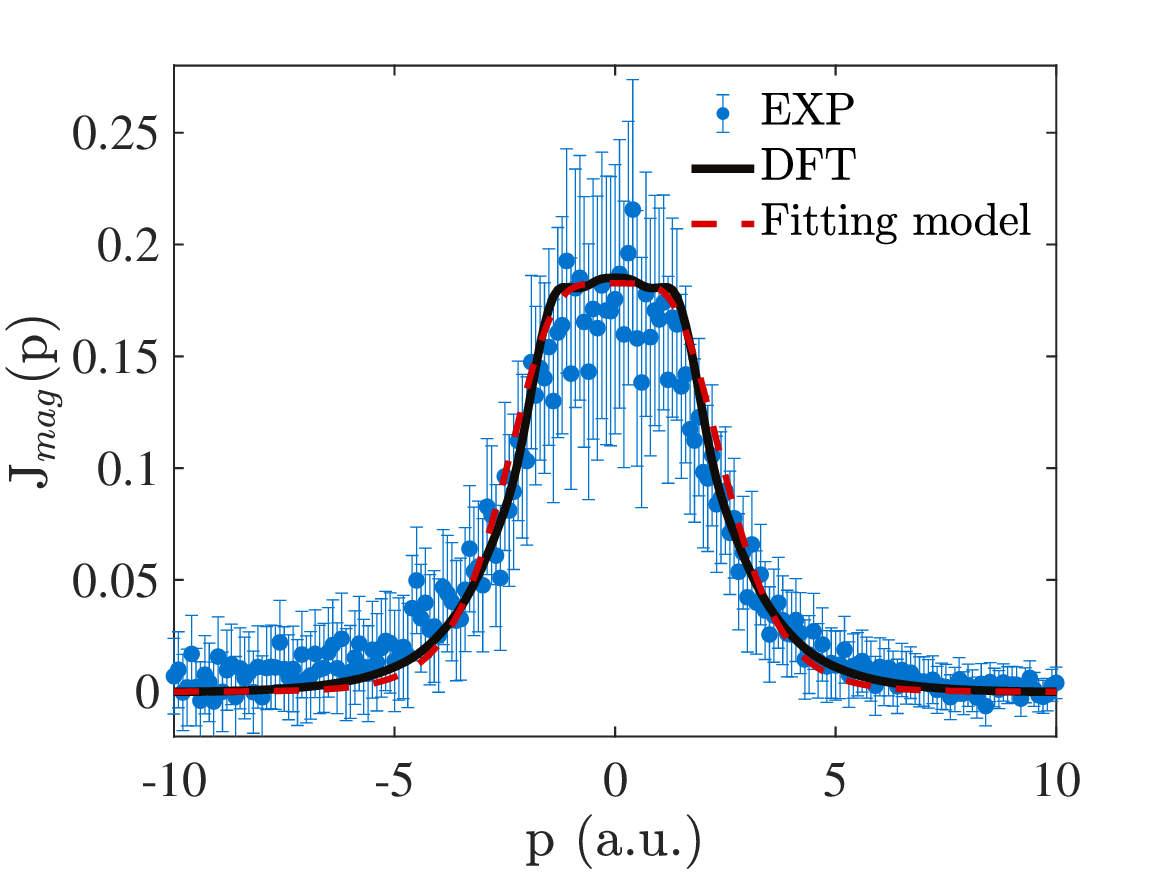}
    \caption{Experimental magnetic Compton profile of Na$_{2/3}$Fe$_{1/2}$Mn$_{1/2}$O$_{2}$ compared with the corresponding  DFT results and two model fits discussed in the text.
    }
    \label{Fig:Mag_Compton}
\end{figure}

\begin{figure}
   \includegraphics[width=.5\linewidth]{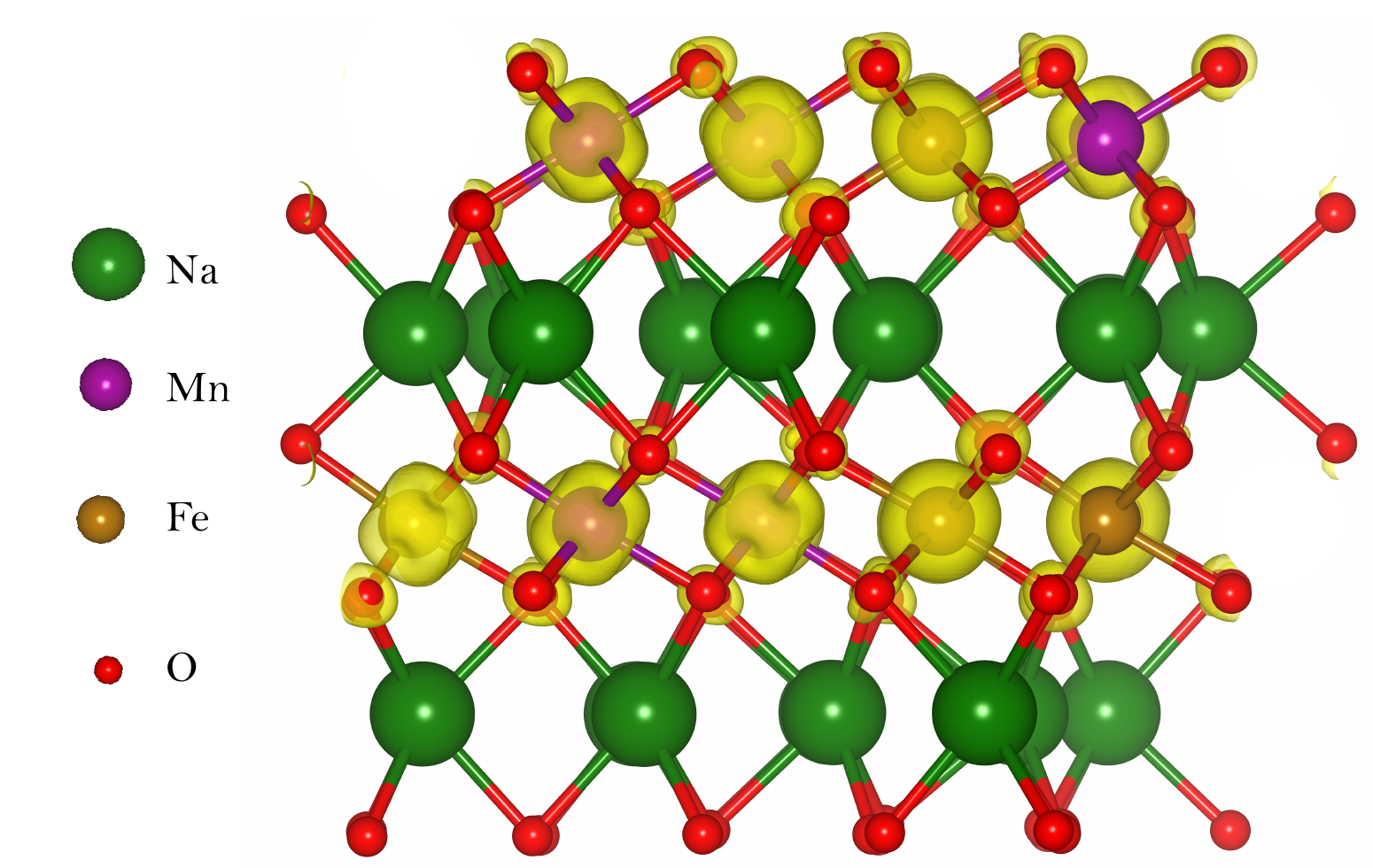}
    \caption{DFT-based spin density in Na$_{2/3}$Fe$_{1/2}$Mn$_{1/2}$O$_{2}$ with an isosurface value of 0.02 a.u. }
    \label{Fig:Spin_density}
\end{figure}

The measured magnetic Compton profile \( J_{\mathrm{mag}} \) shown in Fig.~\ref{Fig:Mag_Compton} indicates that our sample of
Na$_{x}$Fe$_{1/2}$Mn$_{1/2}$O$_{2}$ has a net magnetization at $x=2/3$.
Interestingly, the DFT based magnetic Compton profile for same concentration exhibits remarkable agreement in shape to the measured $x=2/3$ profile when both profiles are normalized to the magnetic moment, as shown in Fig.~\ref{Fig:Mag_Compton}. To rationalize this shape, we use the sum of two Slater terms: a contribution from a transition metal $3d$ Slater orbital with exponent $Z_{\text{TM}} = 3.7$ and another contribution from an O-$2p$ orbital with exponent $Z_{\text{O}} =2$. This fit yields a large O-$2p$ contribution of about 5\,\% obtained by using Chi-square optimization applied to both the experimental noisy curve and the noise-free DFT profile. This magnetic oxygen contribution to $J_{\mathrm{mag}}$ is consistent with the DFT-based magnetization density distribution for $x=2/3$, which is shown in Fig.~\ref{Fig:Spin_density}.
Even though an O$^{2-}$ ion in isolation has a closed 2$p^6$ shell and no net magnetic moment, in the cathode solid-state environment electron holes appear on the oxygen ions, changing their electronic and magnetic behavior.
Interestingly, in our DFT calculations, a significant magnetization is observed from
the oxygen ions in Na$_{2/3}$Fe$_{1/2}$Mn$_{1/2}$O$_2$ with values reaching 0.2\,$\mu_\mathrm{B}$ per O atom.

The magnetic Compton scattering experiment measures a magnetic moment of 0.45~$\mu_\mathrm{B}$ per formula unit for $x = 1/3$ and 0.365~$\mu_\mathrm{B}$ per formula unit for $x = 2/3$.
In comparison,
our DFT calculations that adapt to the ferromagnetic high-spin configuration for Fe and Mn of Zarrabeitia \emph{et al.}~\cite{zarrabeitia2022},
predict much larger spin moments per formula unit: 3.833~$\mu_\mathrm{B}$ for the P2 and 3.167~$\mu_\mathrm{B}$ for the O2 phase.
Note, however, that
Compton scattering measurements capture the total magnetization of the system.
Previous studies~\cite{kim2022} show that the magnetic layers of transition metals are ferromagnetic at low sodium content but become antiferromagnetic with increasing sodium content. This implies that the 2.5~T magnetic field used in our experiments is not strong enough to fully align the spins, which would explain why the measured magnetic moments are smaller than the corresponding theoretical values.
Furthermore,
we find that with the r2SCAN functional, it is possible to stabilize ferrimagnetic states with considerably lower total magnetization.
However, an investigation of these interesting effects is out of the scope of this study.

\begin{figure}
 \includegraphics[width=.5\linewidth]
  {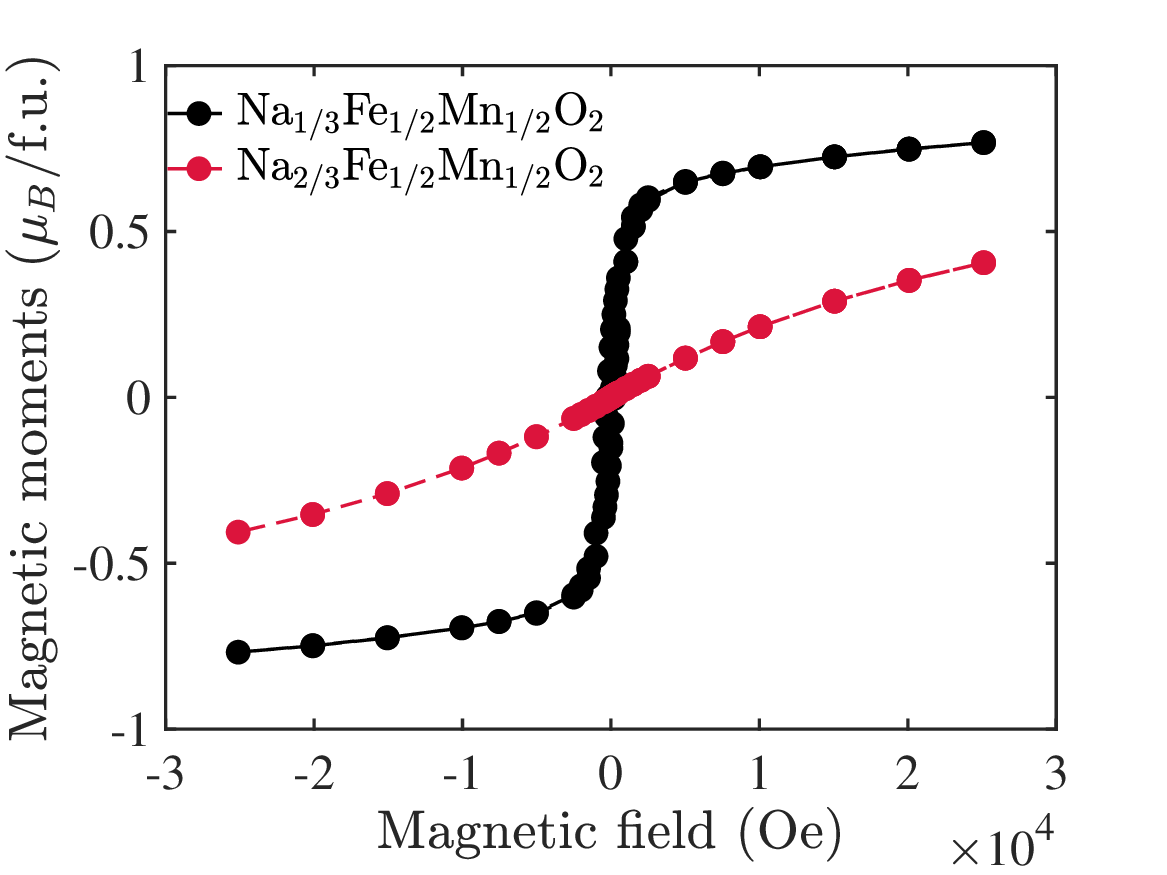}
   \caption{Magnetization curves obtained by SQUID experiments for samples with  $x=1/3$ and $x=2/3$. Measurements were performed at low temperature $T=10$\,K. Magnetic field was scanned from $-2.5$\,T to 2.5\,T.}
\label{Fig:SQUID}
\end{figure}

Fig.~\ref{Fig:SQUID} shows the results of our SQUID measurements. This technique determines the total magnetic moment, which includes both orbital and spin contributions, whereas magnetic Compton scattering probes only the spin moment~\cite{Platzman}. For the $x=2/3$ composition, the total measured magnetic moment is 0.406 \(\mu_\mathrm{B}\). In contrast, the total measured magnetic moment of the $x=1/3$ sample is 0.768 \(\mu_\mathrm{B}\), which is significantly larger than the spin magnetic moment measured by the magnetic Compton scattering experiment. This indicates that the orbital magnetic contribution becomes substantial for $x=1/3$. Finally, our results for $x=2/3$ yield magnetization consistent with SQUID studies by Xu \textit{et al.}~\cite{xu2014}.

\section{Conclusion and Outlook}
We have investigated Na-Fe-Mn-O layered-oxide cathodes for sodium-ion batteries. This stable layered structure enables efficient Na-ion insertion and extraction, delivering promising cycling performance~\cite{yabuuchi2012}. Composed of abundant and eco-friendly elements, this material represents a cost-effective and sustainable alternative to lithium iron phosphate cathodes. Our study demonstrates that x-ray Compton scattering provides a sensitive, bulk-sensitive probe of the electronic structure in Na-ion battery cathodes. Combined total and magnetic Compton scattering measurements, supported by DFT-based momentum-space modeling, provide a detailed orbital and spin-resolved picture of redox activity in Na$_x$Fe$_{1/2}$Mn$_{1/2}$O$_2$ for $x = 1/3$ and $x = 2/3$. We identify oxygen $2p$ states as key contributors to the redox process.

Our analysis reveals the presence of a pronounced negative excursion in the sodiated difference Compton profile $\Delta J(p)$ at higher momenta, reflecting the delocalization of transition-metal $3d$ electrons. This feature, which is associated with electronic conductivity, serves as a descriptor for the sodiated metallic state ($x = 2/3$), and it could guide the optimization of reversible electrochemical devices, where low oxide conductivity limits efficiency. This descriptor is most relevant when O-$2p$ orbitals dominate the redox process, accompanied by a partial delocalization of the transition-metal 3$d$ electrons. We quantified electron transfer between the orbitals and found that extracting Na ions redistributes the $3d$ electrons, leading to an increased hole density in the O-$2p$ orbital. This hole density, in turn, gives rise to the magnetization of the oxygen ion. It would be interesting to independently probe oxygen magnetization via magnetic circular dichroism experiments at the oxygen K-edge ~\cite{Goering2002}.

We emphasize that our DFT r2SCAN calculations do not require any ad hoc Hubbard parameter, making our analysis particularly robust for modeling correlated systems. Moreover, our study demonstrates the value of Compton scattering in gaining an atomic-level understanding of the functional electronic states that govern battery behavior and confirms that Compton scattering is an excellent tool to detect light elements like Oxygen \cite{Veenavee2024c}. This approach provides a promising route for monitoring and optimizing charge transport and redox activity in Na-ion cathodes.

\section*{Acknowledgements}
This research was supported by the INERCOM platform at LUT university.
The authors acknowledge the Finnish CSC-IT Center for Science, Northeastern University's Advanced Scientific Computation Center and the Discovery Cluster for computational resources.
K.S. was supported by JSPS KAKENHI Grant Nos. 22H02103 and 23K23371.
Compton scattering and magnetic Compton scattering experiments were performed with the approval of JASRI (Proposal No. 2023B1498).
Work related to magnetization measurements was conducted at the Institute for Molecular Science, supported by the Advanced Research Infrastructure for Materials and Nanotechnology (JPMXP1224MS1006) of the Ministry of Education, Culture, Sports, Science and Technology (MEXT), Japan.
J.N. acknowledges Emil Aaltonen foundation for financial support.
U.L. acknowledges The Finnish Research Impact Foundation (FRIF) for Tandem Industry Academy Professor funding 2023--2025.
A.A.S.D. and U.L. acknowledge the Big Data project.
L.R. was supported by the Research Council of Finland (decision numbers 359183 and 353095).
The work at Northeastern University was supported by the U.S. Office of Naval Research grant number N00014-23-1-2330 and benefited from the Massachusetts Technology Collaborative award MTC-22032, and the Quantum Materials and Sensing Institute.
We are grateful to Prof. Miguel {\'A}ngel Mu{\~n}oz-M{\'a}rquez for sending us the VASP input files of Ref.~\cite{zarrabeitia2022}.
The authors also acknowledge Erkki Lähderanta, Egor Fadeev, and Andrea Di Feo from LUT University for their valuable insights.
We are grateful to the GENSIB consortium participants, including Jan Kuriplach, Claude Monney, Vincent Garrido, Olli Juntunen,  Minglong He, and Hossein Joya, for their their collaboration.


\begin{thebibliography}{47}%
\makeatletter
\providecommand \@ifxundefined [1]{%
 \@ifx{#1\undefined}
}%
\providecommand \@ifnum [1]{%
 \ifnum #1\expandafter \@firstoftwo
 \else \expandafter \@secondoftwo
 \fi
}%
\providecommand \@ifx [1]{%
 \ifx #1\expandafter \@firstoftwo
 \else \expandafter \@secondoftwo
 \fi
}%
\providecommand \natexlab [1]{#1}%
\providecommand \enquote  [1]{``#1''}%
\providecommand \bibnamefont  [1]{#1}%
\providecommand \bibfnamefont [1]{#1}%
\providecommand \citenamefont [1]{#1}%
\providecommand \href@noop [0]{\@secondoftwo}%
\providecommand \href [0]{\begingroup \@sanitize@url \@href}%
\providecommand \@href[1]{\@@startlink{#1}\@@href}%
\providecommand \@@href[1]{\endgroup#1\@@endlink}%
\providecommand \@sanitize@url [0]{\catcode `\\12\catcode `\$12\catcode
  `\&12\catcode `\#12\catcode `\^12\catcode `\_12\catcode `\%12\relax}%
\providecommand \@@startlink[1]{}%
\providecommand \@@endlink[0]{}%
\providecommand \url  [0]{\begingroup\@sanitize@url \@url }%
\providecommand \@url [1]{\endgroup\@href {#1}{\urlprefix }}%
\providecommand \urlprefix  [0]{URL }%
\providecommand \Eprint [0]{\href }%
\providecommand \doibase [0]{https://doi.org/}%
\providecommand \selectlanguage [0]{\@gobble}%
\providecommand \bibinfo  [0]{\@secondoftwo}%
\providecommand \bibfield  [0]{\@secondoftwo}%
\providecommand \translation [1]{[#1]}%
\providecommand \BibitemOpen [0]{}%
\providecommand \bibitemStop [0]{}%
\providecommand \bibitemNoStop [0]{.\EOS\space}%
\providecommand \EOS [0]{\spacefactor3000\relax}%
\providecommand \BibitemShut  [1]{\csname bibitem#1\endcsname}%
\let\auto@bib@innerbib\@empty
\bibitem [{\citenamefont {Olivetti}\ \emph {et~al.}(2017)\citenamefont
  {Olivetti}, \citenamefont {Ceder}, \citenamefont {Gaustad},\ and\
  \citenamefont {Fu}}]{Elsa2017}%
  \BibitemOpen
  \bibfield  {author} {\bibinfo {author} {\bibfnamefont {E.~A.}\ \bibnamefont
  {Olivetti}}, \bibinfo {author} {\bibfnamefont {G.}~\bibnamefont {Ceder}},
  \bibinfo {author} {\bibfnamefont {G.~G.}\ \bibnamefont {Gaustad}},\ and\
  \bibinfo {author} {\bibfnamefont {X.}~\bibnamefont {Fu}},\ }\bibfield
  {title} {\bibinfo {title} {Lithium-ion battery supply chain considerations:
  Analysis of potential bottlenecks in critical metals},\ }\href
  {https://doi.org/https://doi.org/10.1016/j.joule.2017.08.019} {\bibfield
  {journal} {\bibinfo  {journal} {Joule}\ }\textbf {\bibinfo {volume} {1}},\
  \bibinfo {pages} {229} (\bibinfo {year} {2017})}\BibitemShut {NoStop}%
\bibitem [{\citenamefont {Quinteros-Condoretty}\ \emph
  {et~al.}(2020)\citenamefont {Quinteros-Condoretty}, \citenamefont {Albareda},
  \citenamefont {Barbiellini},\ and\ \citenamefont {Soyer}}]{quinteros2020}%
  \BibitemOpen
  \bibfield  {author} {\bibinfo {author} {\bibfnamefont {A.~R.}\ \bibnamefont
  {Quinteros-Condoretty}}, \bibinfo {author} {\bibfnamefont {L.}~\bibnamefont
  {Albareda}}, \bibinfo {author} {\bibfnamefont {B.}~\bibnamefont
  {Barbiellini}},\ and\ \bibinfo {author} {\bibfnamefont {A.}~\bibnamefont
  {Soyer}},\ }\bibfield  {title} {\bibinfo {title} {A socio-technical
  transition of sustainable lithium industry in latin america},\ }\href@noop {}
  {\bibfield  {journal} {\bibinfo  {journal} {Procedia manufacturing}\ }\textbf
  {\bibinfo {volume} {51}},\ \bibinfo {pages} {1737} (\bibinfo {year}
  {2020})}\BibitemShut {NoStop}%
\bibitem [{\citenamefont {Dorau}\ \emph {et~al.}(2024)\citenamefont {Dorau},
  \citenamefont {Sommer}, \citenamefont {Koloch}, \citenamefont
  {R{\"o}{\ss}-Ohlenroth}, \citenamefont {Schreiber}, \citenamefont {Neuner},
  \citenamefont {Gamra}, \citenamefont {Lin}, \citenamefont {Sch{\"o}berl},
  \citenamefont {Bilfinger} \emph {et~al.}}]{dorau2024}%
  \BibitemOpen
  \bibfield  {author} {\bibinfo {author} {\bibfnamefont {F.~A.}\ \bibnamefont
  {Dorau}}, \bibinfo {author} {\bibfnamefont {A.}~\bibnamefont {Sommer}},
  \bibinfo {author} {\bibfnamefont {J.}~\bibnamefont {Koloch}}, \bibinfo
  {author} {\bibfnamefont {R.}~\bibnamefont {R{\"o}{\ss}-Ohlenroth}}, \bibinfo
  {author} {\bibfnamefont {M.}~\bibnamefont {Schreiber}}, \bibinfo {author}
  {\bibfnamefont {M.}~\bibnamefont {Neuner}}, \bibinfo {author} {\bibfnamefont
  {K.~A.}\ \bibnamefont {Gamra}}, \bibinfo {author} {\bibfnamefont
  {Y.}~\bibnamefont {Lin}}, \bibinfo {author} {\bibfnamefont {J.}~\bibnamefont
  {Sch{\"o}berl}}, \bibinfo {author} {\bibfnamefont {P.}~\bibnamefont
  {Bilfinger}}, \emph {et~al.},\ }\bibfield  {title} {\bibinfo {title}
  {Comprehensive analysis of commercial sodium-ion batteries: Structural and
  electrochemical insights},\ }\href@noop {} {\bibfield  {journal} {\bibinfo
  {journal} {Journal of The Electrochemical Society}\ }\textbf {\bibinfo
  {volume} {171}},\ \bibinfo {pages} {090521} (\bibinfo {year}
  {2024})}\BibitemShut {NoStop}%
\bibitem [{\citenamefont {Jamil}\ \emph {et~al.}(2023)\citenamefont {Jamil},
  \citenamefont {Feng}, \citenamefont {Fasehullah}, \citenamefont {Ali},
  \citenamefont {Wu}, \citenamefont {Guo}, \citenamefont {Jabar}, \citenamefont
  {Mansoor}, \citenamefont {Niu},\ and\ \citenamefont {Xu}}]{jamil2023}%
  \BibitemOpen
  \bibfield  {author} {\bibinfo {author} {\bibfnamefont {S.}~\bibnamefont
  {Jamil}}, \bibinfo {author} {\bibfnamefont {Y.}~\bibnamefont {Feng}},
  \bibinfo {author} {\bibfnamefont {M.}~\bibnamefont {Fasehullah}}, \bibinfo
  {author} {\bibfnamefont {G.}~\bibnamefont {Ali}}, \bibinfo {author}
  {\bibfnamefont {B.}~\bibnamefont {Wu}}, \bibinfo {author} {\bibfnamefont
  {Y.-J.}\ \bibnamefont {Guo}}, \bibinfo {author} {\bibfnamefont
  {B.}~\bibnamefont {Jabar}}, \bibinfo {author} {\bibfnamefont
  {A.}~\bibnamefont {Mansoor}}, \bibinfo {author} {\bibfnamefont {Y.-B.}\
  \bibnamefont {Niu}},\ and\ \bibinfo {author} {\bibfnamefont {M.}~\bibnamefont
  {Xu}},\ }\bibfield  {title} {\bibinfo {title} {Stabilizing anionic redox in
  mn-rich p2-type layered oxide material by mg substitution},\ }\href@noop {}
  {\bibfield  {journal} {\bibinfo  {journal} {Chemical Engineering Journal}\
  }\textbf {\bibinfo {volume} {471}},\ \bibinfo {pages} {144450} (\bibinfo
  {year} {2023})}\BibitemShut {NoStop}%
\bibitem [{\citenamefont {Rostami}\ \emph {et~al.}(2024)\citenamefont
  {Rostami}, \citenamefont {Valio}, \citenamefont {Suominen}, \citenamefont
  {Tynj{\"a}l{\"a}},\ and\ \citenamefont {Lassi}}]{rostami2024}%
  \BibitemOpen
  \bibfield  {author} {\bibinfo {author} {\bibfnamefont {H.}~\bibnamefont
  {Rostami}}, \bibinfo {author} {\bibfnamefont {J.}~\bibnamefont {Valio}},
  \bibinfo {author} {\bibfnamefont {P.}~\bibnamefont {Suominen}}, \bibinfo
  {author} {\bibfnamefont {P.}~\bibnamefont {Tynj{\"a}l{\"a}}},\ and\ \bibinfo
  {author} {\bibfnamefont {U.}~\bibnamefont {Lassi}},\ }\bibfield  {title}
  {\bibinfo {title} {Advancements in cathode technology, recycling strategies,
  and market dynamics: A comprehensive review of sodium ion batteries},\
  }\href@noop {} {\bibfield  {journal} {\bibinfo  {journal} {Chemical
  Engineering Journal}\ ,\ \bibinfo {pages} {153471}} (\bibinfo {year}
  {2024})}\BibitemShut {NoStop}%
\bibitem [{\citenamefont {Delmas}\ \emph {et~al.}(1980)\citenamefont {Delmas},
  \citenamefont {Fouassier},\ and\ \citenamefont {Hagenmuller}}]{delmas19801}%
  \BibitemOpen
  \bibfield  {author} {\bibinfo {author} {\bibfnamefont {C.}~\bibnamefont
  {Delmas}}, \bibinfo {author} {\bibfnamefont {C.}~\bibnamefont {Fouassier}},\
  and\ \bibinfo {author} {\bibfnamefont {P.}~\bibnamefont {Hagenmuller}},\
  }\bibfield  {title} {\bibinfo {title} {Structural classification and
  properties of the layered oxides},\ }\href
  {https://doi.org/https://doi.org/10.1016/0378-4363(80)90214-4} {\bibfield
  {journal} {\bibinfo  {journal} {Physica B+C}\ }\textbf {\bibinfo {volume}
  {99}},\ \bibinfo {pages} {81} (\bibinfo {year} {1980})}\BibitemShut {NoStop}%
\bibitem [{\citenamefont {Yabuuchi}\ \emph {et~al.}(2012)\citenamefont
  {Yabuuchi}, \citenamefont {Kajiyama}, \citenamefont {Iwatate}, \citenamefont
  {Nishikawa}, \citenamefont {Hitomi}, \citenamefont {Okuyama}, \citenamefont
  {Usui}, \citenamefont {Yamada},\ and\ \citenamefont {Komaba}}]{yabuuchi2012}%
  \BibitemOpen
  \bibfield  {author} {\bibinfo {author} {\bibfnamefont {N.}~\bibnamefont
  {Yabuuchi}}, \bibinfo {author} {\bibfnamefont {M.}~\bibnamefont {Kajiyama}},
  \bibinfo {author} {\bibfnamefont {J.}~\bibnamefont {Iwatate}}, \bibinfo
  {author} {\bibfnamefont {H.}~\bibnamefont {Nishikawa}}, \bibinfo {author}
  {\bibfnamefont {S.}~\bibnamefont {Hitomi}}, \bibinfo {author} {\bibfnamefont
  {R.}~\bibnamefont {Okuyama}}, \bibinfo {author} {\bibfnamefont
  {R.}~\bibnamefont {Usui}}, \bibinfo {author} {\bibfnamefont {Y.}~\bibnamefont
  {Yamada}},\ and\ \bibinfo {author} {\bibfnamefont {S.}~\bibnamefont
  {Komaba}},\ }\bibfield  {title} {\bibinfo {title} {P2-type na x [fe1/2mn1/2]
  o2 made from earth-abundant elements for rechargeable na batteries},\
  }\href@noop {} {\bibfield  {journal} {\bibinfo  {journal} {Nature materials}\
  }\textbf {\bibinfo {volume} {11}},\ \bibinfo {pages} {512} (\bibinfo {year}
  {2012})}\BibitemShut {NoStop}%
\bibitem [{\citenamefont {Tang}\ \emph {et~al.}(2024)\citenamefont {Tang},
  \citenamefont {Zhang}, \citenamefont {Zuo}, \citenamefont {Zhou},
  \citenamefont {Zeng}, \citenamefont {Zhang}, \citenamefont {Zhang},
  \citenamefont {Huang}, \citenamefont {Zheng}, \citenamefont {Xu} \emph
  {et~al.}}]{tang2024}%
  \BibitemOpen
  \bibfield  {author} {\bibinfo {author} {\bibfnamefont {Y.}~\bibnamefont
  {Tang}}, \bibinfo {author} {\bibfnamefont {Q.}~\bibnamefont {Zhang}},
  \bibinfo {author} {\bibfnamefont {W.}~\bibnamefont {Zuo}}, \bibinfo {author}
  {\bibfnamefont {S.}~\bibnamefont {Zhou}}, \bibinfo {author} {\bibfnamefont
  {G.}~\bibnamefont {Zeng}}, \bibinfo {author} {\bibfnamefont {B.}~\bibnamefont
  {Zhang}}, \bibinfo {author} {\bibfnamefont {H.}~\bibnamefont {Zhang}},
  \bibinfo {author} {\bibfnamefont {Z.}~\bibnamefont {Huang}}, \bibinfo
  {author} {\bibfnamefont {L.}~\bibnamefont {Zheng}}, \bibinfo {author}
  {\bibfnamefont {J.}~\bibnamefont {Xu}}, \emph {et~al.},\ }\bibfield  {title}
  {\bibinfo {title} {Sustainable layered cathode with suppressed phase
  transition for long-life sodium-ion batteries},\ }\href@noop {} {\bibfield
  {journal} {\bibinfo  {journal} {Nature Sustainability}\ }\textbf {\bibinfo
  {volume} {7}},\ \bibinfo {pages} {348} (\bibinfo {year} {2024})}\BibitemShut
  {NoStop}%
\bibitem [{\citenamefont {Wang}\ \emph {et~al.}(2025)\citenamefont {Wang},
  \citenamefont {Sun}, \citenamefont {Jin}, \citenamefont {Zhao}, \citenamefont
  {Qu}, \citenamefont {Jiao},\ and\ \citenamefont {Liu}}]{wang2025}%
  \BibitemOpen
  \bibfield  {author} {\bibinfo {author} {\bibfnamefont {Y.}~\bibnamefont
  {Wang}}, \bibinfo {author} {\bibfnamefont {Z.}~\bibnamefont {Sun}}, \bibinfo
  {author} {\bibfnamefont {J.}~\bibnamefont {Jin}}, \bibinfo {author}
  {\bibfnamefont {X.}~\bibnamefont {Zhao}}, \bibinfo {author} {\bibfnamefont
  {X.}~\bibnamefont {Qu}}, \bibinfo {author} {\bibfnamefont {L.}~\bibnamefont
  {Jiao}},\ and\ \bibinfo {author} {\bibfnamefont {Y.}~\bibnamefont {Liu}},\
  }\bibfield  {title} {\bibinfo {title} {In situ bulk oxygen vacancy
  manufacturing and surface spinel layer coating enable high-performance na-ion
  layered fe–mn based cathodes},\ }\href
  {https://doi.org/https://doi.org/10.1002/adfm.202504354} {\bibfield
  {journal} {\bibinfo  {journal} {Advanced Functional Materials}\ }\textbf
  {\bibinfo {volume} {n/a}},\ \bibinfo {pages} {2504354} (\bibinfo {year}
  {2025})},\ \Eprint
  {https://arxiv.org/abs/https://advanced.onlinelibrary.wiley.com/doi/pdf/10.1002/adfm.202504354}
  {https://advanced.onlinelibrary.wiley.com/doi/pdf/10.1002/adfm.202504354}
  \BibitemShut {NoStop}%
\bibitem [{\citenamefont {Kothalawala}\ \emph
  {et~al.}(2024{\natexlab{a}})\citenamefont {Kothalawala}, \citenamefont
  {Suzuki}, \citenamefont {Nokelainen}, \citenamefont {Hyv\"onen},
  \citenamefont {Makkonen}, \citenamefont {Barbiellini}, \citenamefont {Hafiz},
  \citenamefont {Tynj\"al\"a}, \citenamefont {Laine}, \citenamefont
  {V\"alikangas}, \citenamefont {Hu}, \citenamefont {Lassi}, \citenamefont
  {Takano}, \citenamefont {Tsuji}, \citenamefont {Amada}, \citenamefont {Devi},
  \citenamefont {Alatalo}, \citenamefont {Sakurai}, \citenamefont {Sakurai},\
  and\ \citenamefont {Bansil}}]{kothalawala2024a}%
  \BibitemOpen
  \bibfield  {author} {\bibinfo {author} {\bibfnamefont {V.~N.}\ \bibnamefont
  {Kothalawala}}, \bibinfo {author} {\bibfnamefont {K.}~\bibnamefont {Suzuki}},
  \bibinfo {author} {\bibfnamefont {J.}~\bibnamefont {Nokelainen}}, \bibinfo
  {author} {\bibfnamefont {A.}~\bibnamefont {Hyv\"onen}}, \bibinfo {author}
  {\bibfnamefont {I.}~\bibnamefont {Makkonen}}, \bibinfo {author}
  {\bibfnamefont {B.}~\bibnamefont {Barbiellini}}, \bibinfo {author}
  {\bibfnamefont {H.}~\bibnamefont {Hafiz}}, \bibinfo {author} {\bibfnamefont
  {P.}~\bibnamefont {Tynj\"al\"a}}, \bibinfo {author} {\bibfnamefont
  {P.}~\bibnamefont {Laine}}, \bibinfo {author} {\bibfnamefont
  {J.}~\bibnamefont {V\"alikangas}}, \bibinfo {author} {\bibfnamefont
  {T.}~\bibnamefont {Hu}}, \bibinfo {author} {\bibfnamefont {U.}~\bibnamefont
  {Lassi}}, \bibinfo {author} {\bibfnamefont {K.}~\bibnamefont {Takano}},
  \bibinfo {author} {\bibfnamefont {N.}~\bibnamefont {Tsuji}}, \bibinfo
  {author} {\bibfnamefont {Y.}~\bibnamefont {Amada}}, \bibinfo {author}
  {\bibfnamefont {A.~A.~S.}\ \bibnamefont {Devi}}, \bibinfo {author}
  {\bibfnamefont {M.}~\bibnamefont {Alatalo}}, \bibinfo {author} {\bibfnamefont
  {Y.}~\bibnamefont {Sakurai}}, \bibinfo {author} {\bibfnamefont
  {H.}~\bibnamefont {Sakurai}},\ and\ \bibinfo {author} {\bibfnamefont
  {A.}~\bibnamefont {Bansil}},\ }\bibfield  {title} {\bibinfo {title} {Compton
  scattering study of strong orbital delocalization in a ${\mathrm{linio}}_{2}$
  cathode},\ }\href {https://doi.org/10.1103/PhysRevB.109.035139} {\bibfield
  {journal} {\bibinfo  {journal} {Phys. Rev. B}\ }\textbf {\bibinfo {volume}
  {109}},\ \bibinfo {pages} {035139} (\bibinfo {year}
  {2024}{\natexlab{a}})}\BibitemShut {NoStop}%
\bibitem [{\citenamefont {Barbiellini}\ \emph {et~al.}(2016)\citenamefont
  {Barbiellini}, \citenamefont {Suzuki}, \citenamefont {Orikasa}, \citenamefont
  {Kaprzyk}, \citenamefont {Itou}, \citenamefont {Yamamoto}, \citenamefont
  {Wang}, \citenamefont {Hafiz}, \citenamefont {Yamada}, \citenamefont
  {Uchimoto}, \citenamefont {Bansil}, \citenamefont {Sakurai},\ and\
  \citenamefont {Sakurai}}]{barbiellini2016}%
  \BibitemOpen
  \bibfield  {author} {\bibinfo {author} {\bibfnamefont {B.}~\bibnamefont
  {Barbiellini}}, \bibinfo {author} {\bibfnamefont {K.}~\bibnamefont {Suzuki}},
  \bibinfo {author} {\bibfnamefont {Y.}~\bibnamefont {Orikasa}}, \bibinfo
  {author} {\bibfnamefont {S.}~\bibnamefont {Kaprzyk}}, \bibinfo {author}
  {\bibfnamefont {M.}~\bibnamefont {Itou}}, \bibinfo {author} {\bibfnamefont
  {K.}~\bibnamefont {Yamamoto}}, \bibinfo {author} {\bibfnamefont {Y.~J.}\
  \bibnamefont {Wang}}, \bibinfo {author} {\bibfnamefont {H.}~\bibnamefont
  {Hafiz}}, \bibinfo {author} {\bibfnamefont {R.}~\bibnamefont {Yamada}},
  \bibinfo {author} {\bibfnamefont {Y.}~\bibnamefont {Uchimoto}}, \bibinfo
  {author} {\bibfnamefont {A.}~\bibnamefont {Bansil}}, \bibinfo {author}
  {\bibfnamefont {Y.}~\bibnamefont {Sakurai}},\ and\ \bibinfo {author}
  {\bibfnamefont {H.}~\bibnamefont {Sakurai}},\ }\bibfield  {title} {\bibinfo
  {title} {Identifying a descriptor for $d$-orbital delocalization in cathodes
  of li batteries based on x-ray compton scattering},\ }\href
  {https://doi.org/10.1063/1.4961055} {\bibfield  {journal} {\bibinfo
  {journal} {Appl. Phys. Lett.}\ }\textbf {\bibinfo {volume} {109}},\ \bibinfo
  {pages} {073102} (\bibinfo {year} {2016})}\BibitemShut {NoStop}%
\bibitem [{\citenamefont {Nokelainen}\ \emph {et~al.}(2022)\citenamefont
  {Nokelainen}, \citenamefont {Barbiellini}, \citenamefont {Kuriplach},
  \citenamefont {Eijt}, \citenamefont {Ferragut}, \citenamefont {Li},
  \citenamefont {Kothalawala}, \citenamefont {Suzuki}, \citenamefont {Sakurai},
  \citenamefont {Hafiz} \emph {et~al.}}]{nokelainen2022}%
  \BibitemOpen
  \bibfield  {author} {\bibinfo {author} {\bibfnamefont {J.}~\bibnamefont
  {Nokelainen}}, \bibinfo {author} {\bibfnamefont {B.}~\bibnamefont
  {Barbiellini}}, \bibinfo {author} {\bibfnamefont {J.}~\bibnamefont
  {Kuriplach}}, \bibinfo {author} {\bibfnamefont {S.}~\bibnamefont {Eijt}},
  \bibinfo {author} {\bibfnamefont {R.}~\bibnamefont {Ferragut}}, \bibinfo
  {author} {\bibfnamefont {X.}~\bibnamefont {Li}}, \bibinfo {author}
  {\bibfnamefont {V.}~\bibnamefont {Kothalawala}}, \bibinfo {author}
  {\bibfnamefont {K.}~\bibnamefont {Suzuki}}, \bibinfo {author} {\bibfnamefont
  {H.}~\bibnamefont {Sakurai}}, \bibinfo {author} {\bibfnamefont
  {H.}~\bibnamefont {Hafiz}}, \emph {et~al.},\ }\bibfield  {title} {\bibinfo
  {title} {Identifying redox orbitals and defects in lithium-ion cathodes with
  compton scattering and positron annihilation spectroscopies: A review},\
  }\href@noop {} {\bibfield  {journal} {\bibinfo  {journal} {Condensed Matter}\
  }\textbf {\bibinfo {volume} {7}},\ \bibinfo {pages} {47} (\bibinfo {year}
  {2022})}\BibitemShut {NoStop}%
\bibitem [{\citenamefont {Zuo}\ \emph {et~al.}(2025)\citenamefont {Zuo},
  \citenamefont {Liu}, \citenamefont {Cai}, \citenamefont {Hu}, \citenamefont
  {Almazrouei}, \citenamefont {Liu}, \citenamefont {Cui}, \citenamefont {Jia},
  \citenamefont {Apodaca}, \citenamefont {Alami}, \citenamefont {Chen},
  \citenamefont {Li}, \citenamefont {Xu}, \citenamefont {Xiao}, \citenamefont
  {Parkinson}, \citenamefont {Yang}, \citenamefont {Xu},\ and\ \citenamefont
  {Amine}}]{khalil2025}%
  \BibitemOpen
  \bibfield  {author} {\bibinfo {author} {\bibfnamefont {W.}~\bibnamefont
  {Zuo}}, \bibinfo {author} {\bibfnamefont {R.}~\bibnamefont {Liu}}, \bibinfo
  {author} {\bibfnamefont {J.}~\bibnamefont {Cai}}, \bibinfo {author}
  {\bibfnamefont {Y.}~\bibnamefont {Hu}}, \bibinfo {author} {\bibfnamefont
  {M.}~\bibnamefont {Almazrouei}}, \bibinfo {author} {\bibfnamefont
  {X.}~\bibnamefont {Liu}}, \bibinfo {author} {\bibfnamefont {T.}~\bibnamefont
  {Cui}}, \bibinfo {author} {\bibfnamefont {X.}~\bibnamefont {Jia}}, \bibinfo
  {author} {\bibfnamefont {E.}~\bibnamefont {Apodaca}}, \bibinfo {author}
  {\bibfnamefont {J.}~\bibnamefont {Alami}}, \bibinfo {author} {\bibfnamefont
  {Z.}~\bibnamefont {Chen}}, \bibinfo {author} {\bibfnamefont {T.}~\bibnamefont
  {Li}}, \bibinfo {author} {\bibfnamefont {W.}~\bibnamefont {Xu}}, \bibinfo
  {author} {\bibfnamefont {X.}~\bibnamefont {Xiao}}, \bibinfo {author}
  {\bibfnamefont {D.}~\bibnamefont {Parkinson}}, \bibinfo {author}
  {\bibfnamefont {Y.}~\bibnamefont {Yang}}, \bibinfo {author} {\bibfnamefont
  {G.-L.}\ \bibnamefont {Xu}},\ and\ \bibinfo {author} {\bibfnamefont
  {K.}~\bibnamefont {Amine}},\ }\bibfield  {title} {\bibinfo {title}
  {Nondestructive analysis of commercial batteries},\ }\href
  {https://doi.org/10.1021/acs.chemrev.4c00566} {\bibfield  {journal} {\bibinfo
   {journal} {Chemical Reviews}\ }\textbf {\bibinfo {volume} {125}},\ \bibinfo
  {pages} {369} (\bibinfo {year} {2025})},\ \bibinfo {note} {pMID: 39688494},\
  \Eprint {https://arxiv.org/abs/https://doi.org/10.1021/acs.chemrev.4c00566}
  {https://doi.org/10.1021/acs.chemrev.4c00566} \BibitemShut {NoStop}%
\bibitem [{\citenamefont {Suzuki}\ \emph {et~al.}(2021)\citenamefont {Suzuki},
  \citenamefont {Suzuki}, \citenamefont {Otsuka}, \citenamefont {Tsuji},
  \citenamefont {Jalkanen}, \citenamefont {Koskinen}, \citenamefont {Hoshi},
  \citenamefont {Honkanen}, \citenamefont {Hafiz}, \citenamefont {Sakurai}
  \emph {et~al.}}]{suzuki2021}%
  \BibitemOpen
  \bibfield  {author} {\bibinfo {author} {\bibfnamefont {K.}~\bibnamefont
  {Suzuki}}, \bibinfo {author} {\bibfnamefont {S.}~\bibnamefont {Suzuki}},
  \bibinfo {author} {\bibfnamefont {Y.}~\bibnamefont {Otsuka}}, \bibinfo
  {author} {\bibfnamefont {N.}~\bibnamefont {Tsuji}}, \bibinfo {author}
  {\bibfnamefont {K.}~\bibnamefont {Jalkanen}}, \bibinfo {author}
  {\bibfnamefont {J.}~\bibnamefont {Koskinen}}, \bibinfo {author}
  {\bibfnamefont {K.}~\bibnamefont {Hoshi}}, \bibinfo {author} {\bibfnamefont
  {A.-P.}\ \bibnamefont {Honkanen}}, \bibinfo {author} {\bibfnamefont
  {H.}~\bibnamefont {Hafiz}}, \bibinfo {author} {\bibfnamefont
  {Y.}~\bibnamefont {Sakurai}}, \emph {et~al.},\ }\bibfield  {title} {\bibinfo
  {title} {Redox oscillations in 18650-type lithium-ion cell revealed by {\it
  in operando} compton scattering imaging},\ }\href
  {https://doi.org/10.1063/5.0048310} {\bibfield  {journal} {\bibinfo
  {journal} {Appl. Phys. Lett.}\ }\textbf {\bibinfo {volume} {118}},\ \bibinfo
  {pages} {161902} (\bibinfo {year} {2021})}\BibitemShut {NoStop}%
\bibitem [{\citenamefont {Suzuki}\ \emph {et~al.}(2024)\citenamefont {Suzuki},
  \citenamefont {Hafiz}, \citenamefont {Kothalawala}, \citenamefont
  {Barbiellini}, \citenamefont {Sakurai},\ and\ \citenamefont
  {Bansil}}]{Suzuki2024}%
  \BibitemOpen
  \bibfield  {author} {\bibinfo {author} {\bibfnamefont {K.}~\bibnamefont
  {Suzuki}}, \bibinfo {author} {\bibfnamefont {H.}~\bibnamefont {Hafiz}},
  \bibinfo {author} {\bibfnamefont {V.~N.}\ \bibnamefont {Kothalawala}},
  \bibinfo {author} {\bibfnamefont {B.}~\bibnamefont {Barbiellini}}, \bibinfo
  {author} {\bibfnamefont {H.}~\bibnamefont {Sakurai}},\ and\ \bibinfo {author}
  {\bibfnamefont {A.}~\bibnamefont {Bansil}},\ }\bibinfo {title} {Rational
  design of battery materials through spectroscopic characterization
  and computational modeling of redox orbitals},\ in\ \href
  {https://doi.org/10.1007/978-3-031-47303-6_22} {\emph {\bibinfo {booktitle}
  {Computational Design of Battery Materials}}},\ \bibinfo {editor} {edited by\
  \bibinfo {editor} {\bibfnamefont {D.~A.~H.}\ \bibnamefont {Hanaor}}}\
  (\bibinfo  {publisher} {Springer International Publishing},\ \bibinfo {year}
  {2024})\ pp.\ \bibinfo {pages} {557--573}\BibitemShut {NoStop}%
\bibitem [{\citenamefont {Leinonen}\ \emph {et~al.}(2025)\citenamefont
  {Leinonen}, \citenamefont {Laine}, \citenamefont {Hu}, \citenamefont
  {Kankaanpää}, \citenamefont {Kervinen}, \citenamefont {Tynjälä},\ and\
  \citenamefont {Lassi}}]{Jere2025}%
  \BibitemOpen
  \bibfield  {author} {\bibinfo {author} {\bibfnamefont {J.}~\bibnamefont
  {Leinonen}}, \bibinfo {author} {\bibfnamefont {P.}~\bibnamefont {Laine}},
  \bibinfo {author} {\bibfnamefont {T.}~\bibnamefont {Hu}}, \bibinfo {author}
  {\bibfnamefont {T.}~\bibnamefont {Kankaanpää}}, \bibinfo {author}
  {\bibfnamefont {I.}~\bibnamefont {Kervinen}}, \bibinfo {author}
  {\bibfnamefont {P.}~\bibnamefont {Tynjälä}},\ and\ \bibinfo {author}
  {\bibfnamefont {U.}~\bibnamefont {Lassi}},\ }\bibfield  {title} {\bibinfo
  {title} {Effect of coprecipitation conditions on the properties of
  fe0.5mn0.5co3 sodium-ion cathode precursors},\ }\href
  {https://doi.org/10.1021/acsomega.5c01869} {\bibfield  {journal} {\bibinfo
  {journal} {ACS Omega}\ }\textbf {\bibinfo {volume} {10}},\ \bibinfo {pages}
  {37128} (\bibinfo {year} {2025})}\BibitemShut {NoStop}%
\bibitem [{\citenamefont {Bl\"ochl}(1994)}]{Blochl1994}%
  \BibitemOpen
  \bibfield  {author} {\bibinfo {author} {\bibfnamefont {P.~E.}\ \bibnamefont
  {Bl\"ochl}},\ }\bibfield  {title} {\bibinfo {title} {Projector augmented-wave
  method},\ }\href {https://doi.org/10.1103/PhysRevB.50.17953} {\bibfield
  {journal} {\bibinfo  {journal} {Phys. Rev. B}\ }\textbf {\bibinfo {volume}
  {50}},\ \bibinfo {pages} {17953} (\bibinfo {year} {1994})}\BibitemShut
  {NoStop}%
\bibitem [{\citenamefont {Kresse}\ and\ \citenamefont
  {Furthmüller}(1996)}]{Kresse1996}%
  \BibitemOpen
  \bibfield  {author} {\bibinfo {author} {\bibfnamefont {G.}~\bibnamefont
  {Kresse}}\ and\ \bibinfo {author} {\bibfnamefont {J.}~\bibnamefont
  {Furthmüller}},\ }\bibfield  {title} {\bibinfo {title} {Efficiency of
  ab-initio total energy calculations for metals and semiconductors using a
  plane-wave basis set},\ }\href {https://doi.org/10.1016/0927-0256(96)00008-0}
  {\bibfield  {journal} {\bibinfo  {journal} {Comput. Mater. Sci.}\ }\textbf
  {\bibinfo {volume} {6}},\ \bibinfo {pages} {1} (\bibinfo {year}
  {1996})}\BibitemShut {NoStop}%
\bibitem [{\citenamefont {Kresse}\ and\ \citenamefont
  {Joubert}(1999)}]{Kresse1999}%
  \BibitemOpen
  \bibfield  {author} {\bibinfo {author} {\bibfnamefont {G.}~\bibnamefont
  {Kresse}}\ and\ \bibinfo {author} {\bibfnamefont {D.}~\bibnamefont
  {Joubert}},\ }\bibfield  {title} {\bibinfo {title} {From ultrasoft
  pseudopotentials to the projector augmented-wave method},\ }\href
  {https://doi.org/10.1103/PhysRevB.59.1758} {\bibfield  {journal} {\bibinfo
  {journal} {Physical review. B}\ }\textbf {\bibinfo {volume} {59}},\ \bibinfo
  {pages} {1758} (\bibinfo {year} {1999})}\BibitemShut {NoStop}%
\bibitem [{\citenamefont {Furness}\ \emph {et~al.}(2020)\citenamefont
  {Furness}, \citenamefont {Kaplan}, \citenamefont {Ning}, \citenamefont
  {Perdew},\ and\ \citenamefont {Sun}}]{Furness2020}%
  \BibitemOpen
  \bibfield  {author} {\bibinfo {author} {\bibfnamefont {J.~W.}\ \bibnamefont
  {Furness}}, \bibinfo {author} {\bibfnamefont {A.~D.}\ \bibnamefont {Kaplan}},
  \bibinfo {author} {\bibfnamefont {J.}~\bibnamefont {Ning}}, \bibinfo {author}
  {\bibfnamefont {J.~P.}\ \bibnamefont {Perdew}},\ and\ \bibinfo {author}
  {\bibfnamefont {J.}~\bibnamefont {Sun}},\ }\bibfield  {title} {\bibinfo
  {title} {Accurate and numerically efficient r2scan meta-generalized gradient
  approximation},\ }\href {https://doi.org/10.1021/acs.jpclett.0c02405}
  {\bibfield  {journal} {\bibinfo  {journal} {The Journal of Physical Chemistry
  Letters}\ }\textbf {\bibinfo {volume} {11}},\ \bibinfo {pages} {8208}
  (\bibinfo {year} {2020})},\ \bibinfo {note} {pMID: 32876454},\ \Eprint
  {https://arxiv.org/abs/https://doi.org/10.1021/acs.jpclett.0c02405}
  {https://doi.org/10.1021/acs.jpclett.0c02405} \BibitemShut {NoStop}%
\bibitem [{\citenamefont {Dudarev}\ \emph {et~al.}(1998)\citenamefont
  {Dudarev}, \citenamefont {Botton}, \citenamefont {Savrasov}, \citenamefont
  {Humphreys},\ and\ \citenamefont {Sutton}}]{dudarev1998}%
  \BibitemOpen
  \bibfield  {author} {\bibinfo {author} {\bibfnamefont {S.~L.}\ \bibnamefont
  {Dudarev}}, \bibinfo {author} {\bibfnamefont {G.~A.}\ \bibnamefont {Botton}},
  \bibinfo {author} {\bibfnamefont {S.~Y.}\ \bibnamefont {Savrasov}}, \bibinfo
  {author} {\bibfnamefont {C.~J.}\ \bibnamefont {Humphreys}},\ and\ \bibinfo
  {author} {\bibfnamefont {A.~P.}\ \bibnamefont {Sutton}},\ }\bibfield  {title}
  {\bibinfo {title} {Electron-energy-loss spectra and the structural stability
  of nickel oxide: An lsda+u study},\ }\href
  {https://doi.org/10.1103/PhysRevB.57.1505} {\bibfield  {journal} {\bibinfo
  {journal} {Phys. Rev. B}\ }\textbf {\bibinfo {volume} {57}},\ \bibinfo
  {pages} {1505} (\bibinfo {year} {1998})}\BibitemShut {NoStop}%
\bibitem [{\citenamefont {Lee}\ \emph {et~al.}(2013)\citenamefont {Lee},
  \citenamefont {Xu},\ and\ \citenamefont {Meng}}]{shirley2013}%
  \BibitemOpen
  \bibfield  {author} {\bibinfo {author} {\bibfnamefont {D.~H.}\ \bibnamefont
  {Lee}}, \bibinfo {author} {\bibfnamefont {J.}~\bibnamefont {Xu}},\ and\
  \bibinfo {author} {\bibfnamefont {Y.~S.}\ \bibnamefont {Meng}},\ }\bibfield
  {title} {\bibinfo {title} {An advanced cathode for na-ion batteries with high
  rate and excellent structural stability},\ }\href
  {https://doi.org/10.1039/C2CP44467D} {\bibfield  {journal} {\bibinfo
  {journal} {Phys. Chem. Chem. Phys.}\ }\textbf {\bibinfo {volume} {15}},\
  \bibinfo {pages} {3304} (\bibinfo {year} {2013})}\BibitemShut {NoStop}%
\bibitem [{\citenamefont {Zarrabeitia}\ \emph {et~al.}(2022)\citenamefont
  {Zarrabeitia}, \citenamefont {Nobili}, \citenamefont {Lakuntza},
  \citenamefont {Carrasco}, \citenamefont {Rojo}, \citenamefont
  {Casas-Cabanas},\ and\ \citenamefont
  {Mu{\~n}oz-M{\'a}rquez}}]{zarrabeitia2022}%
  \BibitemOpen
  \bibfield  {author} {\bibinfo {author} {\bibfnamefont {M.}~\bibnamefont
  {Zarrabeitia}}, \bibinfo {author} {\bibfnamefont {F.}~\bibnamefont {Nobili}},
  \bibinfo {author} {\bibfnamefont {O.}~\bibnamefont {Lakuntza}}, \bibinfo
  {author} {\bibfnamefont {J.}~\bibnamefont {Carrasco}}, \bibinfo {author}
  {\bibfnamefont {T.}~\bibnamefont {Rojo}}, \bibinfo {author} {\bibfnamefont
  {M.}~\bibnamefont {Casas-Cabanas}},\ and\ \bibinfo {author} {\bibfnamefont
  {M.~{\'A}.}\ \bibnamefont {Mu{\~n}oz-M{\'a}rquez}},\ }\bibfield  {title}
  {\bibinfo {title} {Role of the voltage window on the capacity retention of
  p2-na2/3 [fe1/2mn1/2] o2 cathode material for rechargeable sodium-ion
  batteries},\ }\href@noop {} {\bibfield  {journal} {\bibinfo  {journal}
  {Communications chemistry}\ }\textbf {\bibinfo {volume} {5}},\ \bibinfo
  {pages} {11} (\bibinfo {year} {2022})}\BibitemShut {NoStop}%
\bibitem [{\citenamefont {Mortemard~de Boisse}\ \emph
  {et~al.}(2014)\citenamefont {Mortemard~de Boisse}, \citenamefont {Carlier},
  \citenamefont {Guignard}, \citenamefont {Bourgeois},\ and\ \citenamefont
  {Delmas}}]{delmas2014}%
  \BibitemOpen
  \bibfield  {author} {\bibinfo {author} {\bibfnamefont {B.}~\bibnamefont
  {Mortemard~de Boisse}}, \bibinfo {author} {\bibfnamefont {D.}~\bibnamefont
  {Carlier}}, \bibinfo {author} {\bibfnamefont {M.}~\bibnamefont {Guignard}},
  \bibinfo {author} {\bibfnamefont {L.}~\bibnamefont {Bourgeois}},\ and\
  \bibinfo {author} {\bibfnamefont {C.}~\bibnamefont {Delmas}},\ }\bibfield
  {title} {\bibinfo {title} {P2-naxmn1/2fe1/2o2 phase used as positive
  electrode in na batteries: Structural changes induced by the electrochemical
  (de)intercalation process},\ }\href {https://doi.org/10.1021/ic5017802}
  {\bibfield  {journal} {\bibinfo  {journal} {Inorganic Chemistry}\ }\textbf
  {\bibinfo {volume} {53}},\ \bibinfo {pages} {11197} (\bibinfo {year}
  {2014})},\ \bibinfo {note} {pMID: 25255369},\ \Eprint
  {https://arxiv.org/abs/https://doi.org/10.1021/ic5017802}
  {https://doi.org/10.1021/ic5017802} \BibitemShut {NoStop}%
\bibitem [{\citenamefont {Tang}\ \emph {et~al.}(2009)\citenamefont {Tang},
  \citenamefont {Sanville},\ and\ \citenamefont {Henkelman}}]{2009_Tang_Bader}%
  \BibitemOpen
  \bibfield  {author} {\bibinfo {author} {\bibfnamefont {W.}~\bibnamefont
  {Tang}}, \bibinfo {author} {\bibfnamefont {E.}~\bibnamefont {Sanville}},\
  and\ \bibinfo {author} {\bibfnamefont {G.}~\bibnamefont {Henkelman}},\
  }\bibfield  {title} {\bibinfo {title} {A grid-based bader analysis algorithm
  without lattice bias},\ }\href
  {https://doi.org/10.1088/0953-8984/21/8/084204} {\bibfield  {journal}
  {\bibinfo  {journal} {J. Phys.: Condens. Matter}\ }\textbf {\bibinfo {volume}
  {21}},\ \bibinfo {pages} {084204} (\bibinfo {year} {2009})}\BibitemShut
  {NoStop}%
\bibitem [{\citenamefont {Yu}\ and\ \citenamefont
  {Trinkle}(2011)}]{2011_Yu_Trinkle_Bader_improvement}%
  \BibitemOpen
  \bibfield  {author} {\bibinfo {author} {\bibfnamefont {M.}~\bibnamefont
  {Yu}}\ and\ \bibinfo {author} {\bibfnamefont {D.~R.}\ \bibnamefont
  {Trinkle}},\ }\bibfield  {title} {\bibinfo {title} {Accurate and efficient
  algorithm for bader charge integration},\ }\href
  {https://doi.org/10.1063/1.3553716} {\bibfield  {journal} {\bibinfo
  {journal} {J. Chem. Phys.}\ }\textbf {\bibinfo {volume} {134}},\ \bibinfo
  {pages} {064111} (\bibinfo {year} {2011})}\BibitemShut {NoStop}%
\bibitem [{\citenamefont {Cooper}\ \emph {et~al.}(2004)\citenamefont {Cooper},
  \citenamefont {Mijnarends}, \citenamefont {Shiotani}, \citenamefont {Sakai},\
  and\ \citenamefont {Bansil}}]{Cooper2004}%
  \BibitemOpen
  \bibfield  {author} {\bibinfo {author} {\bibfnamefont {M.}~\bibnamefont
  {Cooper}}, \bibinfo {author} {\bibfnamefont {P.}~\bibnamefont {Mijnarends}},
  \bibinfo {author} {\bibfnamefont {N.}~\bibnamefont {Shiotani}}, \bibinfo
  {author} {\bibfnamefont {N.}~\bibnamefont {Sakai}},\ and\ \bibinfo {author}
  {\bibfnamefont {A.}~\bibnamefont {Bansil}},\ }\bibinfo {title} {X-ray compton
  scattering}\ (\bibinfo  {publisher} {Oxford University Press},\ \bibinfo
  {address} {Oxford},\ \bibinfo {year} {2004})\ pp.\ \bibinfo {pages}
  {31--39}\BibitemShut {NoStop}%
\bibitem [{\citenamefont {Barbiellini}\ and\ \citenamefont
  {Bansil}(2001)}]{barbiellini2001}%
  \BibitemOpen
  \bibfield  {author} {\bibinfo {author} {\bibfnamefont {B.}~\bibnamefont
  {Barbiellini}}\ and\ \bibinfo {author} {\bibfnamefont {A.}~\bibnamefont
  {Bansil}},\ }\bibfield  {title} {\bibinfo {title} {Treatment of correlation
  effects in electron momentum density: Density functional theory and beyond},\
  }\href@noop {} {\bibfield  {journal} {\bibinfo  {journal} {Journal of Physics
  and Chemistry of Solids}\ }\textbf {\bibinfo {volume} {62}},\ \bibinfo
  {pages} {2181} (\bibinfo {year} {2001})}\BibitemShut {NoStop}%
\bibitem [{\citenamefont {Kaplan}\ \emph {et~al.}(2003)\citenamefont {Kaplan},
  \citenamefont {Barbiellini},\ and\ \citenamefont {Bansil}}]{Kaplan2003}%
  \BibitemOpen
  \bibfield  {author} {\bibinfo {author} {\bibfnamefont {I.~G.}\ \bibnamefont
  {Kaplan}}, \bibinfo {author} {\bibfnamefont {B.}~\bibnamefont
  {Barbiellini}},\ and\ \bibinfo {author} {\bibfnamefont {A.}~\bibnamefont
  {Bansil}},\ }\bibfield  {title} {\bibinfo {title} {Compton scattering beyond
  the impulse approximation},\ }\href
  {https://doi.org/10.1103/PhysRevB.68.235104} {\bibfield  {journal} {\bibinfo
  {journal} {Phys. Rev. B}\ }\textbf {\bibinfo {volume} {68}},\ \bibinfo
  {pages} {235104} (\bibinfo {year} {2003})}\BibitemShut {NoStop}%
\bibitem [{\citenamefont {Barbiellini}\ and\ \citenamefont
  {Bansil}(2024)}]{barbiellini2024}%
  \BibitemOpen
  \bibfield  {author} {\bibinfo {author} {\bibfnamefont {B.}~\bibnamefont
  {Barbiellini}}\ and\ \bibinfo {author} {\bibfnamefont {A.}~\bibnamefont
  {Bansil}},\ }\bibfield  {title} {\bibinfo {title} {{Scattering techniques,
  Compton}},\ }in\ \href
  {https://doi.org/https://doi.org/10.1016/B978-0-323-90800-9.00107-4} {\emph
  {\bibinfo {booktitle} {Encyclopedia of Condensed Matter Physics (Second
  Edition)}}},\ \bibinfo {editor} {edited by\ \bibinfo {editor} {\bibfnamefont
  {T.}~\bibnamefont {Chakraborty}}}\ (\bibinfo  {publisher} {Academic Press},\
  \bibinfo {address} {Oxford},\ \bibinfo {year} {2024})\ \bibinfo {edition}
  {2nd}\ ed.,\ pp.\ \bibinfo {pages} {173--186}\BibitemShut {NoStop}%
\bibitem [{\citenamefont {Biggs}\ \emph {et~al.}(1975)\citenamefont {Biggs},
  \citenamefont {Mendelsohn},\ and\ \citenamefont {Mann}}]{biggs1975}%
  \BibitemOpen
  \bibfield  {author} {\bibinfo {author} {\bibfnamefont {F.}~\bibnamefont
  {Biggs}}, \bibinfo {author} {\bibfnamefont {L.}~\bibnamefont {Mendelsohn}},\
  and\ \bibinfo {author} {\bibfnamefont {J.}~\bibnamefont {Mann}},\ }\bibfield
  {title} {\bibinfo {title} {Hartree-fock compton profiles for the elements},\
  }\href@noop {} {\bibfield  {journal} {\bibinfo  {journal} {Atomic data and
  nuclear data tables}\ }\textbf {\bibinfo {volume} {16}},\ \bibinfo {pages}
  {201} (\bibinfo {year} {1975})}\BibitemShut {NoStop}%
\bibitem [{\citenamefont {Suzuki}\ \emph
  {et~al.}(2016{\natexlab{a}})\citenamefont {Suzuki}, \citenamefont
  {Barbiellini}, \citenamefont {Orikasa}, \citenamefont {Kaprzyk},
  \citenamefont {Itou}, \citenamefont {Yamamoto}, \citenamefont {Wang},
  \citenamefont {Hafiz}, \citenamefont {Uchimoto}, \citenamefont {Bansil} \emph
  {et~al.}}]{suzuki2016}%
  \BibitemOpen
  \bibfield  {author} {\bibinfo {author} {\bibfnamefont {K.}~\bibnamefont
  {Suzuki}}, \bibinfo {author} {\bibfnamefont {B.}~\bibnamefont {Barbiellini}},
  \bibinfo {author} {\bibfnamefont {Y.}~\bibnamefont {Orikasa}}, \bibinfo
  {author} {\bibfnamefont {S.}~\bibnamefont {Kaprzyk}}, \bibinfo {author}
  {\bibfnamefont {M.}~\bibnamefont {Itou}}, \bibinfo {author} {\bibfnamefont
  {K.}~\bibnamefont {Yamamoto}}, \bibinfo {author} {\bibfnamefont {Y.~J.}\
  \bibnamefont {Wang}}, \bibinfo {author} {\bibfnamefont {H.}~\bibnamefont
  {Hafiz}}, \bibinfo {author} {\bibfnamefont {Y.}~\bibnamefont {Uchimoto}},
  \bibinfo {author} {\bibfnamefont {A.}~\bibnamefont {Bansil}}, \emph
  {et~al.},\ }\bibfield  {title} {\bibinfo {title} {Non-destructive measurement
  of in-operando lithium concentration in batteries via x-ray compton
  scattering},\ }\href@noop {} {\bibfield  {journal} {\bibinfo  {journal}
  {Journal of Applied Physics}\ }\textbf {\bibinfo {volume} {119}} (\bibinfo
  {year} {2016}{\natexlab{a}})}\BibitemShut {NoStop}%
\bibitem [{\citenamefont {Makkonen}\ \emph {et~al.}(2005)\citenamefont
  {Makkonen}, \citenamefont {Hakala},\ and\ \citenamefont
  {Puska}}]{Makkonen2005}%
  \BibitemOpen
  \bibfield  {author} {\bibinfo {author} {\bibfnamefont {I.}~\bibnamefont
  {Makkonen}}, \bibinfo {author} {\bibfnamefont {M.}~\bibnamefont {Hakala}},\
  and\ \bibinfo {author} {\bibfnamefont {M.}~\bibnamefont {Puska}},\ }\bibfield
   {title} {\bibinfo {title} {Calculation of valence electron momentum
  densities using the projector augmented-wave method},\ }\href
  {https://doi.org/https://doi.org/10.1016/j.jpcs.2005.02.009} {\bibfield
  {journal} {\bibinfo  {journal} {Journal of Physics and Chemistry of Solids}\
  }\textbf {\bibinfo {volume} {66}},\ \bibinfo {pages} {1128} (\bibinfo {year}
  {2005})}\BibitemShut {NoStop}%
\bibitem [{\citenamefont {Sakurai}(1998)}]{Sakurai1998}%
  \BibitemOpen
  \bibfield  {author} {\bibinfo {author} {\bibfnamefont {Y.}~\bibnamefont
  {Sakurai}},\ }\bibfield  {title} {\bibinfo {title} {{High-Energy
  Inelastic-Scattering Beamline for Electron Momentum Density Study}},\ }\href
  {https://doi.org/10.1107/S0909049598002052} {\bibfield  {journal} {\bibinfo
  {journal} {Journal of Synchrotron Radiation}\ }\textbf {\bibinfo {volume}
  {5}},\ \bibinfo {pages} {208} (\bibinfo {year} {1998})}\BibitemShut {NoStop}%
\bibitem [{\citenamefont {Kakutani}\ \emph {et~al.}(2003)\citenamefont
  {Kakutani}, \citenamefont {Kubo}, \citenamefont {Koizumi}, \citenamefont
  {Sakai}, \citenamefont {Ahuja},\ and\ \citenamefont {Sharma}}]{Kakutani2003}%
  \BibitemOpen
  \bibfield  {author} {\bibinfo {author} {\bibfnamefont {Y.}~\bibnamefont
  {Kakutani}}, \bibinfo {author} {\bibfnamefont {Y.}~\bibnamefont {Kubo}},
  \bibinfo {author} {\bibfnamefont {A.}~\bibnamefont {Koizumi}}, \bibinfo
  {author} {\bibfnamefont {N.}~\bibnamefont {Sakai}}, \bibinfo {author}
  {\bibfnamefont {B.~L.}\ \bibnamefont {Ahuja}},\ and\ \bibinfo {author}
  {\bibfnamefont {B.~K.}\ \bibnamefont {Sharma}},\ }\bibfield  {title}
  {\bibinfo {title} {Magnetic compton profiles of fcc-ni, fcc-fe50ni50 and
  hcp-co},\ }\href {https://doi.org/10.1143/JPSJ.72.599} {\bibfield  {journal}
  {\bibinfo  {journal} {Journal of the Physical Society of Japan}\ }\textbf
  {\bibinfo {volume} {72}},\ \bibinfo {pages} {599} (\bibinfo {year}
  {2003})}\BibitemShut {NoStop}%
\bibitem [{\citenamefont {Perdew}\ \emph {et~al.}(1996)\citenamefont {Perdew},
  \citenamefont {Burke},\ and\ \citenamefont {Ernzerhof}}]{Perdew1996}%
  \BibitemOpen
  \bibfield  {author} {\bibinfo {author} {\bibfnamefont {J.}~\bibnamefont
  {Perdew}}, \bibinfo {author} {\bibfnamefont {K.}~\bibnamefont {Burke}},\ and\
  \bibinfo {author} {\bibfnamefont {M.}~\bibnamefont {Ernzerhof}},\ }\bibfield
  {title} {\bibinfo {title} {Generalized gradient approximation made simple},\
  }\href {https://doi.org/10.1103/PhysRevLett.77.3865} {\bibfield  {journal}
  {\bibinfo  {journal} {Phys. Rev. Lett.}\ }\textbf {\bibinfo {volume} {77}},\
  \bibinfo {pages} {3865} (\bibinfo {year} {1996})}\BibitemShut {NoStop}%
\bibitem [{\citenamefont {Abate}\ \emph {et~al.}(2021)\citenamefont {Abate},
  \citenamefont {Kim}, \citenamefont {Pemmaraju}, \citenamefont {Toney},
  \citenamefont {Yang}, \citenamefont {Devereaux}, \citenamefont {Chueh},\ and\
  \citenamefont {Nazar}}]{abate2021}%
  \BibitemOpen
  \bibfield  {author} {\bibinfo {author} {\bibfnamefont {I.}~\bibnamefont
  {Abate}}, \bibinfo {author} {\bibfnamefont {S.~Y.}\ \bibnamefont {Kim}},
  \bibinfo {author} {\bibfnamefont {C.~D.}\ \bibnamefont {Pemmaraju}}, \bibinfo
  {author} {\bibfnamefont {M.~F.}\ \bibnamefont {Toney}}, \bibinfo {author}
  {\bibfnamefont {W.}~\bibnamefont {Yang}}, \bibinfo {author} {\bibfnamefont
  {T.~P.}\ \bibnamefont {Devereaux}}, \bibinfo {author} {\bibfnamefont {W.~C.}\
  \bibnamefont {Chueh}},\ and\ \bibinfo {author} {\bibfnamefont {L.~F.}\
  \bibnamefont {Nazar}},\ }\bibfield  {title} {\bibinfo {title} {The role of
  metal substitution in tuning anion redox in sodium metal layered oxides
  revealed by x-ray spectroscopy and theory},\ }\href@noop {} {\bibfield
  {journal} {\bibinfo  {journal} {Angewandte Chemie International Edition}\
  }\textbf {\bibinfo {volume} {60}},\ \bibinfo {pages} {10880} (\bibinfo {year}
  {2021})}\BibitemShut {NoStop}%
\bibitem [{\citenamefont {Kim}\ \emph {et~al.}(2022)\citenamefont {Kim},
  \citenamefont {Kim}, \citenamefont {Cho},\ and\ \citenamefont
  {Kim}}]{kim2022}%
  \BibitemOpen
  \bibfield  {author} {\bibinfo {author} {\bibfnamefont {M.}~\bibnamefont
  {Kim}}, \bibinfo {author} {\bibfnamefont {H.}~\bibnamefont {Kim}}, \bibinfo
  {author} {\bibfnamefont {M.}~\bibnamefont {Cho}},\ and\ \bibinfo {author}
  {\bibfnamefont {D.}~\bibnamefont {Kim}},\ }\bibfield  {title} {\bibinfo
  {title} {Theoretical understanding of oxygen stability in mn--fe binary
  layered oxides for sodium-ion batteries},\ }\href@noop {} {\bibfield
  {journal} {\bibinfo  {journal} {Journal of Materials Chemistry A}\ }\textbf
  {\bibinfo {volume} {10}},\ \bibinfo {pages} {11101} (\bibinfo {year}
  {2022})}\BibitemShut {NoStop}%
\bibitem [{\citenamefont {Patel}\ \emph {et~al.}(2022)\citenamefont {Patel},
  \citenamefont {Guruswamy}, \citenamefont {Krzysko}, \citenamefont
  {Charalambous}, \citenamefont {Gades}, \citenamefont {Wiaderek},
  \citenamefont {Quaranta}, \citenamefont {Ren}, \citenamefont {Yakovenko},
  \citenamefont {Ruett} \emph {et~al.}}]{patel2022}%
  \BibitemOpen
  \bibfield  {author} {\bibinfo {author} {\bibfnamefont {U.}~\bibnamefont
  {Patel}}, \bibinfo {author} {\bibfnamefont {T.}~\bibnamefont {Guruswamy}},
  \bibinfo {author} {\bibfnamefont {A.}~\bibnamefont {Krzysko}}, \bibinfo
  {author} {\bibfnamefont {H.}~\bibnamefont {Charalambous}}, \bibinfo {author}
  {\bibfnamefont {L.}~\bibnamefont {Gades}}, \bibinfo {author} {\bibfnamefont
  {K.}~\bibnamefont {Wiaderek}}, \bibinfo {author} {\bibfnamefont
  {O.}~\bibnamefont {Quaranta}}, \bibinfo {author} {\bibfnamefont
  {Y.}~\bibnamefont {Ren}}, \bibinfo {author} {\bibfnamefont {A.}~\bibnamefont
  {Yakovenko}}, \bibinfo {author} {\bibfnamefont {U.}~\bibnamefont {Ruett}},
  \emph {et~al.},\ }\bibfield  {title} {\bibinfo {title} {High-resolution
  compton spectroscopy using x-ray microcalorimeters},\ }\href@noop {}
  {\bibfield  {journal} {\bibinfo  {journal} {Review of Scientific
  Instruments}\ }\textbf {\bibinfo {volume} {93}},\ \bibinfo {pages} {113105}
  (\bibinfo {year} {2022})}\BibitemShut {NoStop}%
\bibitem [{\citenamefont {Kothalawala}\ \emph
  {et~al.}(2024{\natexlab{b}})\citenamefont {Kothalawala}, \citenamefont
  {Suzuki}, \citenamefont {Li}, \citenamefont {Barbiellini}, \citenamefont
  {Nokelainen}, \citenamefont {Makkonen}, \citenamefont {Ferragut},
  \citenamefont {Tynj{\"a}l{\"a}}, \citenamefont {Laine}, \citenamefont
  {V{\"a}likangas} \emph {et~al.}}]{kothalawala2024b}%
  \BibitemOpen
  \bibfield  {author} {\bibinfo {author} {\bibfnamefont {V.~N.}\ \bibnamefont
  {Kothalawala}}, \bibinfo {author} {\bibfnamefont {K.}~\bibnamefont {Suzuki}},
  \bibinfo {author} {\bibfnamefont {X.}~\bibnamefont {Li}}, \bibinfo {author}
  {\bibfnamefont {B.}~\bibnamefont {Barbiellini}}, \bibinfo {author}
  {\bibfnamefont {J.}~\bibnamefont {Nokelainen}}, \bibinfo {author}
  {\bibfnamefont {I.}~\bibnamefont {Makkonen}}, \bibinfo {author}
  {\bibfnamefont {R.}~\bibnamefont {Ferragut}}, \bibinfo {author}
  {\bibfnamefont {P.}~\bibnamefont {Tynj{\"a}l{\"a}}}, \bibinfo {author}
  {\bibfnamefont {P.}~\bibnamefont {Laine}}, \bibinfo {author} {\bibfnamefont
  {J.}~\bibnamefont {V{\"a}likangas}}, \emph {et~al.},\ }\bibfield  {title}
  {\bibinfo {title} {Determining effects of doping lithium nickel oxide with
  tungsten using compton scattering},\ }\href@noop {} {\bibfield  {journal}
  {\bibinfo  {journal} {APL Energy}\ }\textbf {\bibinfo {volume} {2}},\
  \bibinfo {pages} {026102} (\bibinfo {year} {2024}{\natexlab{b}})}\BibitemShut
  {NoStop}%
\bibitem [{\citenamefont {Suzuki}\ \emph
  {et~al.}(2016{\natexlab{b}})\citenamefont {Suzuki}, \citenamefont
  {Barbiellini}, \citenamefont {Orikasa}, \citenamefont {Kaprzyk},
  \citenamefont {Itou}, \citenamefont {Yamamoto}, \citenamefont {Wang},
  \citenamefont {Hafiz}, \citenamefont {Uchimoto}, \citenamefont {Bansil},
  \citenamefont {Sakurai},\ and\ \citenamefont {Sakurai}}]{Kosuke2016}%
  \BibitemOpen
  \bibfield  {author} {\bibinfo {author} {\bibfnamefont {K.}~\bibnamefont
  {Suzuki}}, \bibinfo {author} {\bibfnamefont {B.}~\bibnamefont {Barbiellini}},
  \bibinfo {author} {\bibfnamefont {Y.}~\bibnamefont {Orikasa}}, \bibinfo
  {author} {\bibfnamefont {S.}~\bibnamefont {Kaprzyk}}, \bibinfo {author}
  {\bibfnamefont {M.}~\bibnamefont {Itou}}, \bibinfo {author} {\bibfnamefont
  {K.}~\bibnamefont {Yamamoto}}, \bibinfo {author} {\bibfnamefont {Y.~J.}\
  \bibnamefont {Wang}}, \bibinfo {author} {\bibfnamefont {H.}~\bibnamefont
  {Hafiz}}, \bibinfo {author} {\bibfnamefont {Y.}~\bibnamefont {Uchimoto}},
  \bibinfo {author} {\bibfnamefont {A.}~\bibnamefont {Bansil}}, \bibinfo
  {author} {\bibfnamefont {Y.}~\bibnamefont {Sakurai}},\ and\ \bibinfo {author}
  {\bibfnamefont {H.}~\bibnamefont {Sakurai}},\ }\bibfield  {title} {\bibinfo
  {title} {Non-destructive measurement of in-operando lithium concentration in
  batteries via x-ray compton scattering},\ }\href
  {https://doi.org/10.1063/1.4939304} {\bibfield  {journal} {\bibinfo
  {journal} {Journal of Applied Physics}\ }\textbf {\bibinfo {volume} {119}},\
  \bibinfo {pages} {025103} (\bibinfo {year} {2016}{\natexlab{b}})}\BibitemShut
  {NoStop}%
\bibitem [{\citenamefont {Suzuki}\ \emph {et~al.}(2015)\citenamefont {Suzuki},
  \citenamefont {Barbiellini}, \citenamefont {Orikasa}, \citenamefont {Go},
  \citenamefont {Sakurai}, \citenamefont {Kaprzyk}, \citenamefont {Itou},
  \citenamefont {Yamamoto}, \citenamefont {Uchimoto}, \citenamefont {Wang},
  \citenamefont {Hafiz}, \citenamefont {Bansil},\ and\ \citenamefont
  {Sakurai}}]{suzuki2015}%
  \BibitemOpen
  \bibfield  {author} {\bibinfo {author} {\bibfnamefont {K.}~\bibnamefont
  {Suzuki}}, \bibinfo {author} {\bibfnamefont {B.}~\bibnamefont {Barbiellini}},
  \bibinfo {author} {\bibfnamefont {Y.}~\bibnamefont {Orikasa}}, \bibinfo
  {author} {\bibfnamefont {N.}~\bibnamefont {Go}}, \bibinfo {author}
  {\bibfnamefont {H.}~\bibnamefont {Sakurai}}, \bibinfo {author} {\bibfnamefont
  {S.}~\bibnamefont {Kaprzyk}}, \bibinfo {author} {\bibfnamefont
  {M.}~\bibnamefont {Itou}}, \bibinfo {author} {\bibfnamefont {K.}~\bibnamefont
  {Yamamoto}}, \bibinfo {author} {\bibfnamefont {Y.}~\bibnamefont {Uchimoto}},
  \bibinfo {author} {\bibfnamefont {Y.~J.}\ \bibnamefont {Wang}}, \bibinfo
  {author} {\bibfnamefont {H.}~\bibnamefont {Hafiz}}, \bibinfo {author}
  {\bibfnamefont {A.}~\bibnamefont {Bansil}},\ and\ \bibinfo {author}
  {\bibfnamefont {Y.}~\bibnamefont {Sakurai}},\ }\bibfield  {title} {\bibinfo
  {title} {Extracting the redox orbitals in li battery materials with
  high-resolution x-ray compton scattering spectroscopy},\ }\href
  {https://doi.org/10.1103/PhysRevLett.114.087401} {\bibfield  {journal}
  {\bibinfo  {journal} {Phys. Rev. Lett.}\ }\textbf {\bibinfo {volume} {114}},\
  \bibinfo {pages} {087401} (\bibinfo {year} {2015})}\BibitemShut {NoStop}%
\bibitem [{\citenamefont {Hafiz}\ \emph {et~al.}(2021)\citenamefont {Hafiz},
  \citenamefont {Suzuki}, \citenamefont {Barbiellini}, \citenamefont {Tsuji},
  \citenamefont {Yabuuchi}, \citenamefont {Yamamoto}, \citenamefont {Orikasa},
  \citenamefont {Uchimoto}, \citenamefont {Sakurai}, \citenamefont {Sakurai}
  \emph {et~al.}}]{hafiz2021}%
  \BibitemOpen
  \bibfield  {author} {\bibinfo {author} {\bibfnamefont {H.}~\bibnamefont
  {Hafiz}}, \bibinfo {author} {\bibfnamefont {K.}~\bibnamefont {Suzuki}},
  \bibinfo {author} {\bibfnamefont {B.}~\bibnamefont {Barbiellini}}, \bibinfo
  {author} {\bibfnamefont {N.}~\bibnamefont {Tsuji}}, \bibinfo {author}
  {\bibfnamefont {N.}~\bibnamefont {Yabuuchi}}, \bibinfo {author}
  {\bibfnamefont {K.}~\bibnamefont {Yamamoto}}, \bibinfo {author}
  {\bibfnamefont {Y.}~\bibnamefont {Orikasa}}, \bibinfo {author} {\bibfnamefont
  {Y.}~\bibnamefont {Uchimoto}}, \bibinfo {author} {\bibfnamefont
  {Y.}~\bibnamefont {Sakurai}}, \bibinfo {author} {\bibfnamefont
  {H.}~\bibnamefont {Sakurai}}, \emph {et~al.},\ }\bibfield  {title} {\bibinfo
  {title} {Tomographic reconstruction of oxygen orbitals in lithium-rich
  battery materials},\ }\href {https://doi.org/10.1038/s41586-021-03509-z}
  {\bibfield  {journal} {\bibinfo  {journal} {Nature}\ }\textbf {\bibinfo
  {volume} {594}},\ \bibinfo {pages} {213} (\bibinfo {year}
  {2021})}\BibitemShut {NoStop}%
\bibitem [{\citenamefont {Platzman}\ and\ \citenamefont
  {Tzoar}(1970)}]{Platzman}%
  \BibitemOpen
  \bibfield  {author} {\bibinfo {author} {\bibfnamefont {P.~M.}\ \bibnamefont
  {Platzman}}\ and\ \bibinfo {author} {\bibfnamefont {N.}~\bibnamefont
  {Tzoar}},\ }\bibfield  {title} {\bibinfo {title} {Magnetic scattering of x
  rays from electrons in molecules and solids},\ }\href
  {https://doi.org/10.1103/PhysRevB.2.3556} {\bibfield  {journal} {\bibinfo
  {journal} {Phys. Rev. B}\ }\textbf {\bibinfo {volume} {2}},\ \bibinfo {pages}
  {3556} (\bibinfo {year} {1970})}\BibitemShut {NoStop}%
\bibitem [{\citenamefont {Xu}\ \emph {et~al.}(2014)\citenamefont {Xu},
  \citenamefont {Chou}, \citenamefont {Wang}, \citenamefont {Liu},\ and\
  \citenamefont {Dou}}]{xu2014}%
  \BibitemOpen
  \bibfield  {author} {\bibinfo {author} {\bibfnamefont {J.}~\bibnamefont
  {Xu}}, \bibinfo {author} {\bibfnamefont {S.-L.}\ \bibnamefont {Chou}},
  \bibinfo {author} {\bibfnamefont {J.-L.}\ \bibnamefont {Wang}}, \bibinfo
  {author} {\bibfnamefont {H.-K.}\ \bibnamefont {Liu}},\ and\ \bibinfo {author}
  {\bibfnamefont {S.-X.}\ \bibnamefont {Dou}},\ }\bibfield  {title} {\bibinfo
  {title} {Layered p2-na0.66fe0.5mn0.5o2 cathode material for rechargeable
  sodium-ion batteries},\ }\href
  {https://doi.org/https://doi.org/10.1002/celc.201300026} {\bibfield
  {journal} {\bibinfo  {journal} {ChemElectroChem}\ }\textbf {\bibinfo {volume}
  {1}},\ \bibinfo {pages} {371} (\bibinfo {year} {2014})},\ \Eprint
  {https://arxiv.org/abs/https://chemistry-europe.onlinelibrary.wiley.com/doi/pdf/10.1002/celc.201300026}
  {https://chemistry-europe.onlinelibrary.wiley.com/doi/pdf/10.1002/celc.201300026}
  \BibitemShut {NoStop}%
\bibitem [{\citenamefont {Goering}\ \emph {et~al.}(2002)\citenamefont
  {Goering}, \citenamefont {Bayer}, \citenamefont {Gold}, \citenamefont
  {Schütz}, \citenamefont {Rabe}, \citenamefont {Rüdiger},\ and\
  \citenamefont {Güntherodt}}]{Goering2002}%
  \BibitemOpen
  \bibfield  {author} {\bibinfo {author} {\bibfnamefont {E.}~\bibnamefont
  {Goering}}, \bibinfo {author} {\bibfnamefont {A.}~\bibnamefont {Bayer}},
  \bibinfo {author} {\bibfnamefont {S.}~\bibnamefont {Gold}}, \bibinfo {author}
  {\bibfnamefont {G.}~\bibnamefont {Schütz}}, \bibinfo {author} {\bibfnamefont
  {M.}~\bibnamefont {Rabe}}, \bibinfo {author} {\bibfnamefont {U.}~\bibnamefont
  {Rüdiger}},\ and\ \bibinfo {author} {\bibfnamefont {G.}~\bibnamefont
  {Güntherodt}},\ }\bibfield  {title} {\bibinfo {title} {Direct correlation of
  cr 3d orbital polarization and o k-edge x-ray magnetic circular dichroism of
  epitaxial cro2 films},\ }\href {https://doi.org/10.1209/epl/i2002-00103-6}
  {\bibfield  {journal} {\bibinfo  {journal} {Europhysics Letters}\ }\textbf
  {\bibinfo {volume} {58}},\ \bibinfo {pages} {906} (\bibinfo {year}
  {2002})}\BibitemShut {NoStop}%
\bibitem [{\citenamefont {Kothalawala}\ \emph
  {et~al.}(2024{\natexlab{c}})\citenamefont {Kothalawala}, \citenamefont
  {Guruswamy}, \citenamefont {Quaranta}, \citenamefont {Patel}, \citenamefont
  {Gades}, \citenamefont {Taddei}, \citenamefont {Yakovenko}, \citenamefont
  {Zheng}, \citenamefont {Morgan}, \citenamefont {Weber}, \citenamefont {Yan},
  \citenamefont {Swetz}, \citenamefont {Makkonen}, \citenamefont {Yeddu},
  \citenamefont {Bansil}, \citenamefont {Ruett}, \citenamefont {Miceli},
  \citenamefont {Nokelainen},\ and\ \citenamefont
  {Barbiellini}}]{Veenavee2024c}%
  \BibitemOpen
  \bibfield  {author} {\bibinfo {author} {\bibfnamefont {V.~N.}\ \bibnamefont
  {Kothalawala}}, \bibinfo {author} {\bibfnamefont {T.}~\bibnamefont
  {Guruswamy}}, \bibinfo {author} {\bibfnamefont {O.}~\bibnamefont {Quaranta}},
  \bibinfo {author} {\bibfnamefont {U.~M.}\ \bibnamefont {Patel}}, \bibinfo
  {author} {\bibfnamefont {L.}~\bibnamefont {Gades}}, \bibinfo {author}
  {\bibfnamefont {K.}~\bibnamefont {Taddei}}, \bibinfo {author} {\bibfnamefont
  {A.}~\bibnamefont {Yakovenko}}, \bibinfo {author} {\bibfnamefont
  {M.}~\bibnamefont {Zheng}}, \bibinfo {author} {\bibfnamefont
  {K.}~\bibnamefont {Morgan}}, \bibinfo {author} {\bibfnamefont
  {J.}~\bibnamefont {Weber}}, \bibinfo {author} {\bibfnamefont
  {D.}~\bibnamefont {Yan}}, \bibinfo {author} {\bibfnamefont {D.}~\bibnamefont
  {Swetz}}, \bibinfo {author} {\bibfnamefont {I.}~\bibnamefont {Makkonen}},
  \bibinfo {author} {\bibfnamefont {H.~K.}\ \bibnamefont {Yeddu}}, \bibinfo
  {author} {\bibfnamefont {A.}~\bibnamefont {Bansil}}, \bibinfo {author}
  {\bibfnamefont {U.}~\bibnamefont {Ruett}}, \bibinfo {author} {\bibfnamefont
  {A.}~\bibnamefont {Miceli}}, \bibinfo {author} {\bibfnamefont
  {J.}~\bibnamefont {Nokelainen}},\ and\ \bibinfo {author} {\bibfnamefont
  {B.}~\bibnamefont {Barbiellini}},\ }\bibfield  {title} {\bibinfo {title}
  {Extracting the electronic structure of light elements in bulk materials
  through a compton scattering method in the readily accessible hard x-ray
  regime},\ }\href {https://doi.org/10.1063/5.0207375} {\bibfield  {journal}
  {\bibinfo  {journal} {Applied Physics Letters}\ }\textbf {\bibinfo {volume}
  {124}},\ \bibinfo {pages} {223501} (\bibinfo {year}
  {2024}{\natexlab{c}})}\BibitemShut {NoStop}%
\end{thebibliography}

\providecommand{\noopsort}[1]{}\providecommand{\singleletter}[1]{#1}%

\end{document}